\documentclass[12pt]{article}
\usepackage[dvips]{graphicx}

\def\lromn#1{\uppercase\expandafter{\romannumeral#1}}

\begin{document}

\begin{center}
\begin{large}
\textbf{
Neutrino Pair Emission from Excited Atoms
}

\end{large}
\end{center}

\vspace{2cm}
\begin{center}
\begin{large}
M. Yoshimura

Center of Quantum Universe and
Department of Physics, Okayama University, \\
Tsushima-naka 3-1-1, Okayama 700-8530,
Japan
\end{large}
\end{center}

\vspace{2cm}
\begin{center}
\begin{Large}
{\bf ABSTRACT}
\end{Large}
\end{center}

We explore a possibility of measuring the absolute magnitude and the 
nature (Majorana vs Dirac) of neutrino masses, by using a novel process of
neutrino pair emission from metastable excited atoms.
Except lepton number non-conserving processes, 
the neutrino pair ($\nu \bar{\nu}$)
emission is the unique process to directly distinguish the Majorana neutrino from
the Dirac neutrino, using the interference effect of identical fermions.
The small energy difference between atomic levels makes it easier
to measure small neutrino masses 
as indicated by neutrino oscillation experiments.
The crucial point is how to enhance the rate of pair emission
without enhancing the radiative decay.
We discuss two particular cases;
(1) laser irradiated pair emission from metastable atoms, and (2)
microwave irradiated emission from circular Rydberg states.
A new mechanism of the parametric amplification to enhance the
neutrino pair emission is pointed out when Rydberg atoms are irradiated by
microwave, while the radiative process may be inhibited
by the cavity QED effect.
A great variety of measurable neutrino parameters and a variety
of experimental methods make this investigation attractive.

\vspace{2cm}

\section{Introduction}

\vspace{0.5cm} 
\hspace*{0.5cm} 
The nature of neutrino masses, along with its precise values
and their mixing parameters which appear 
in the weak interaction, is of fundamental
importance to explore physics far beyond the standard model.
In particular, whether the neutrino belongs to
the special class of neutral particles described by
the Majorana equation, or to the usual Dirac particle
we are so familiar with, is a central issue of great interest.

In the present work we propose a novel approach to answer this
important issue, the Majorana vs the Dirac particle,
and suggest new experimental methods to do this.
Moreover, we would like to suggest a method
to simultaneously determine 
absolute values of neutrino masses in the same experiment.

The neutrino masses indicated by recent oscillation
experiments suggest, but do not determine the hierarchical mass 
pattern.
One tends to take a view that two mass scales suggested by
the atmospheric neutrino and the solar neutrino oscillation
is close to two heavier neutrino masses, with a small correction
from the lightest neutrino;
\begin{eqnarray}
&&
m_3 \sim 50 meV \,, \hspace{0.3cm}
m_2 \sim 10 meV \,, \hspace{0.3cm}
m_1 \ll m_2 \,.
\end{eqnarray}
This is the case of the normal hierarchy. On the other hand,
in the case of the inverted hierarchy one has the mass relation;
\(\,
m_3 \approx m_2 \sim 50 meV \,, m_3^2 - m_2^2 \sim (10 meV)^2
\,, m_1 \ll m_2 \,.
\,\)
How small the lightest mass $m_1$ is and
how much the heaviest neutrino of mass $m_3$ 
is mixed in the flavor state $\nu_e$ 
are both important questions unanswered by neutrino oscillation
experiments so far.
It is desirable for a single, well-organized experiment to 
be able to address all these questions.
Indeed, our proposal directly attempts to answer all these
problems with an extra bonus;
if the method works ideally, one may hope to 
embark on the neutrino mass spectroscopy, along with
determination of the Majorana vs Dirac particle.

Available energy difference between atomic levels is closest 
to small neutrino masses indicated by neutrino oscillation.
Other energy scales are much larger; for instance the tritium
beta endpoint $\sim 18.6 keV$.
Among others, Rydberg states \cite{rydberg atom}
of a large principal quantum number $n$
have energy difference to nearest levels of order,
\begin{eqnarray}
&&
\Delta E \sim
27 \, meV \Delta n (\frac{n}{10})^{-3}
\,.
\end{eqnarray}
This makes it urgent to seriously consider atomic experiments for
the neutrino mass measurement, if the rate lies within the
experimental reach.
The pair emission rate scales with $G_F^2 E^5$ 
with the energy $E$, 
the constant
being the Fermi coupling $G_F \sim 10^{-5} GeV^{-2}$, hence it is
usually impossible to have a reasonable rate, unless
some novel mechanism of enhancement is proposed.

In the present work we discuss two different types of enhanced atomic transitions;
\(\,
\gamma + I^* \rightarrow I^{**} + \nu_i \nu_j
\,,
\,\)
where $\gamma$ is either a laser photon or a microwave photon.
The inital atomic state $I^*$ is a metastable excited state of a long
lifetime, for instance $> 1 sec$, while
the final state $I^{**}$ has a strong E1 rate to a lower level such as
the ground state.
The experimental signal would be a detection of transition to the level
$I^{**}$ experimentally designed vacant initially.
For unambiguous identification of a weak interaction process
such as this neutrino pair emission it is desirable to
measure a parity violating quantity such as the rate difference
between initial different circular polarizations.
In the first case of laser irradiated pair emission, 
one uses a resonance effect for enhancement.
Our second, and a more ambitious proposal is to utilize 
an inherent instability and
its associated enhanced decay of neutrino pair emission
when Rydberg states are irradiated by a strong microwave field, 
at the same time using the principle of inhibition of ordinary radiative
decay in a microwave cavity, a cavity QED effect \cite{inhibited emission}.

The laser irradiated pair emission, perhaps using a more conventional
experimental technique,
might be a shortest route towards
establishing the largest mass $m_3$ and the distinction of Majorana and Dirac
neutrino.
The Rydberg atom may lead to a more complete neutrino spectroscopy,
including a precision determination of smaller masses and mixing angles.

To the best of our knowledge the neutrino pair emission from
atomic excited states, either
spontaneous or photon initiated, has not been observed so far, or even not
discussed extensively in the literature, presumably due to a clear
lack of interest.
We wish to point out here that the atomic pair emission is ideal
for a precision neutrino spectroscopy.

\vspace{0.5cm}

The rest of the paper is organized as follows.

In section 1 we describe in detail how to 
distinguish the Majorana and the Dirac neutrinos.
Our approach uses the 2-component formalism for both
cases of the Majorana and the Dirac fields.
Neutrinos that participate in the standard weak interaction
of the $SU(3) \times SU(2) \times U(1)$ gauge theory are
described by chirally projected 2-component spinors.
We find it most unambiguous and straightforward to use the 2-component
spinor both for neutrinos and electrons under the nuclear
Coulomb field, in order to clarify and unambiguously identify
the true nature of massive Majorana neutrino.
This is done using a representation of $4\times 4$ gamma matrices
as given in Appendix A,
which is not the most popular one
in the literature, but is explained in many textbooks such as
\cite{aithinson-hey}.
A comparison of the 2- and the 4-component approaches 
is also given in ref. \cite{aithinson-hey}.

The most popular approach uses the 4-component spinor $\psi$ 
with the Majorana condition $\psi^C = \psi$, which essentially
reduces the 4-component $\psi$ to the 2-component spinor,
$\varphi \sim \psi^C + \psi$.
A great merit of the 2-component formalism is that
it uses independent variables alone.
On the other hand, the 4-component formalism uses
redundant fields constrained by the Majorana condtion.
We prefer to use independent components alone since the neutrino
that appears in the usual weak interaction needs two components 
$\varphi$ alone.
In our 2-component approach it is made evident
below that the distinction of the massive Majorana and
the Dirac cases occurs only via the interference term
of two identical particles present in the Majorana case.

We then present the pair $\nu \bar{\nu}$ emission
amplitude and demonstrate how the distinction
arises in the two cases.
The other place where the Majorana nature arises
might be in lepton number non-conserving processes such as
the neutrinoless bouble beta decay.
But in the case of lepton number non-conservation
there can be no proof that the Majorana neutrino
is directly involved, since there might be another source
of lepton number non-conservation..

In section 2 we work out kinematical
factors of the pair emission from excited atoms.
In section 3 we discuss one of the enhanced
process; laser irradiated pair emission.
How the threshold behavior of photon irradiated
pair emission differs in the Majorana
and the Dirac neutrinos is described in detail.
We then numerically estimate the rate for this process
assuming a standard laser flux of commercially available frequency
resolution.
We shall further illustrate how to determine
neutrino parameters including 3 mass values and
the angle $\theta_{13}$ \cite{cp measurement method} 
if the proposed experiment becomes possible.

In section 4 an entirely new
process of pair emission from circular Rydberg
atoms is discussed.
Both the standard multiphoton picture and more
general effect of the parametric amplification
is described.
The enhancement factor of the microwave irradiation
is interpreted in the multiphoton picure
as the existence of a great many varieties of paths of
stimulated photon emission that
bridge between the initial and the final states
of energy difference of order of the sum of the mass
of two emitted neutrinos.
The multiphoton process corresponds to the narrow
band region of the parametric resonance.
What is more interesting is the wide band region 
of the parametric amplification
which is missing in the multiphoton picture, and
one may expect an even larger, exponential growth of the rate.
We discuss the unitarity bound on the pair emission rate
in the wide band region.

Our basic assumption throughout this paper is that the standard
electroweak theory correctly describes the neutrino interaction,
while their small masses, either of the Majorana or of the
Dirac type, are generated at a much higher energy
scale than the electroweak symmetry breaking.
Thus, the weak current associated with neutrinos is always 
taken of the $V - A$ form.
We warn that introduction of $V + A$ current may drastically
change the result presented here.

Throughout this paper we use the natural unit of
$\hbar = 1$ and $c=1$, and
$\alpha \sim 1/137$ and $\alpha^2 m_e/2 \sim 13.6 eV$.

\vspace{1cm}
\section{How to distinguish the Majorana neutrino from the
Dirac neutrino}

\vspace{0.5cm} 
\hspace*{0.5cm}
The most commonly assumed method of observing the Majorana
nature of neutrino masses is to discover the lepton number
non-conservation, as typically exemplified by the neutrinoless
double beta decay.
But, this is not the unique way of detection.
Another, and a more direct method is to exploit the identical particle
effect of Majorana particles \cite{kpi}, since a Majorana particle
is identical to its anti-particle.

In the non-relativistic regime where the distinction of
Majorana vs Dirac particles is expected to appear,
plane-wave solutions for the Majorana and the 
Dirac particles might appear different, and this difference
might show up in many places if momentum of neutrinos is
smaller than their masses.
We need a systematic way to handle the most general cases.
This is the purpose of the present section.
In the end we shall show that
there exists a representation demonstrating
wave functions common to both the Majorana and the Dirac cases.
All other representations should give equivalent results to
this one.

Since a convenient account of the Majorana field, in particular
of the 2-component formalism, is missing
\cite{aithinson-hey}, 
we explain in detail and give fundamental formulas related to Majorana neutrinos,
in particular an explicit plane-wave solution for the massive
Majorana particle.
Our approach is based on the 2-component spinor,
and unlike many other works,
in no place we adopt the 4-component description.
This way we believe that a possible complication due to
a constrained fermion field of 4-component description is avoided.
We shall prove that the interference term 
of the anti-symmetrized wave function of identical fermions
is the only source of distinction of the Majorana and the Dirac
neutrinos.
A great merit of our 2-component approach is that
this simple result is an automatic consequence of
our formalism.

Fine details are relegated to Appendix A, some of which
should be useful in different contexts.

\vspace{0.5cm}
\subsection{Majorana equation}

\hspace*{0.5cm}
Lorentz invariance allows electrically neutral particles
to be described by two component spinor equation, as
pointed out a long time ago by Majorana.
The Majorana equation for free neutrinos is 
\begin{eqnarray}
&&
(i\partial_t - i\vec{\sigma}\cdot\vec{\nabla}) \varphi = i m 
\sigma_2 \varphi^*
\,,
\label{majorana equation}
\end{eqnarray}
with $m$ the neutrino mass.
The 2-component spinor $\varphi$ belongs to an irreducible representation
of the Lorentz group, unlike reducible representation of
the 4-component Dirac spinor.
The most salient feature of this equation is that
it contains $\varphi$ as well as its conjugate $\varphi^*$, thus
the lepton number is violated, or more properly,
one cannot define the lepton number.
Unless the lepton number is violated in other places of interaction,
the rate of lepton number violating processes is
proportional to the square of the neutrino mass,
actually some weighted average of neutrino masses squared.

The plane-wave solution to eq.(\ref{majorana equation}) is given by
\begin{eqnarray}
&&
\varphi_p(x) = 
e^{-i p\cdot x} 
\left(
\begin{array}{c}
a  \\
b    
\end{array}
\right)
+ 
e^{i p\cdot x} 
\left(
\begin{array}{c}
 c \\
 d   
\end{array}
\right)
\,,
\\ &&
\left(
\begin{array}{c}
c  \\
d 
\end{array}
\right)
=
\frac{E_p - \vec{\sigma}\cdot\vec{p}}{m}
(-i \sigma_2)
\left(
\begin{array}{c}
a^*  \\
b^*
\end{array}
\right)
\,.
\end{eqnarray}
Consistent quantization as discussed in Appendix A
leads to the normalized operator
form of plane-wave solution written in terms of the helicity
eigenstate of eigenvalue $h$,
\begin{eqnarray}
&&
\hspace*{-1cm}
\varphi_{\vec{p}\,,h}(x) = c(\vec{p}\,,h)
e^{-ip\cdot x}u(\vec{p}\,, h)
+ c^{\dagger}(\vec{p}\,, - h)e^{ip\cdot x} 
\sqrt{\frac{E_p + hp }{E_p - hp}}(-i \sigma_2) u^*(\vec{p}\,, h)
\,,
\label{anti-particle creation}
\\ &&
u(\vec{p}\,, h)
=
\frac{1}{2}
\sqrt{\frac{E_p - hp}{pE_p(p + hp_3)}}
\left(
\begin{array}{c}
  p + hp_3 \\
  h(p_1 + ip_2)
\end{array}
\right)
\,.
\label{m-spinor}
\end{eqnarray}
The helicity eigenstate wave fucntion satisfys
$(\vec{\sigma}\cdot\vec{p}/p) u(\vec{p}\,, h) = h  u(\vec{p}\,, h)$.
Quantization of the Majorana field as explained in Appendix A
gives the interpretation of 
$c(\vec{p}\,,h)$ and $c^{\dagger}(\vec{p}\,, -h)
= (\,c(\vec{p}\,, h)\,)^{\dagger}$
as annihilation and creation operators of Majorana particles 
of momentum $\vec{p}$ and helicity $h$.

\vspace{0.5cm}
\subsection{Weak interaction of neutrino}

\hspace*{0.5cm}
We only consider weak interaction of neutrinos with electron, since
our subject is the atomic weak process, hence we
ignore heavier charged leptons and quarks.

The Majorana neutrino field appears only in the form of 
the projected 2-component spinor,
$\varphi = (1 - \gamma_5)\psi/2 $ in all weak processes.
We shall also write down the electron field operator decomposed
into the 2-component form,
which must be done using the same representation 
of $\gamma$ matrices as done for the neutrino.

It is convenient to use the Fierz transformed 4-Fermi form
including both charged current (CC) and neutral current (NC) interactions;
\begin{eqnarray}
&& 
\frac{G_F}{\sqrt{2}}\,
\bar{\nu}^e \gamma_{\alpha}(1 - \gamma_5)\nu^e
\bar{e}\gamma^{\alpha}(1 - \gamma_5) e
\nonumber \\ &&
- \frac{G_F}{2\sqrt{2}}\,
\sum_i \bar{\nu}^i \gamma_{\alpha}(1 - \gamma_5)\nu^i \bar{e}
\left(
\gamma^{\alpha}
(1 - 4\sin^2 \theta_W - \gamma_5) 
\right) e
\,,
\label{weak off e}
\end{eqnarray}
where $\sin^2 \theta_W \approx 0.231$ experimentally.
The relative sign of CC and NC terms becomes important
later.

A care must be taken of the effect of the nuclear Coulomb field
on electrons. Fortunately, to orders of $\alpha$ and $1/m_e$,
the result is simple since $(E - m_e - V)/m_e = O[\alpha^2]$ 
terms can be neglected.
The result of non-relativistic limit is summarized as
\begin{eqnarray}
&&
2\sqrt{2}G_F\times \,
\left[
\left(
\nu_e^{\dagger} \nu_e e^{\dagger} e
+ \nu_e^{\dagger} \vec{\sigma} \nu_e \cdot e^{\dagger}\vec{\sigma} e
\right)
\right.
\nonumber \\ &&
- \frac{1}{2} \sum_i
\left(
\nu_i^{\dagger} \nu_i e^{\dagger} \left[
1-  4\sin^2 \theta_W ( 1 + \frac{i}{m_e}\vec{\sigma}\cdot
\vec{\nabla}) \right] e
\right.
\nonumber \\ &&
\left.
\left.
+ \nu_i^{\dagger} \vec{\sigma} \nu_i \cdot e^{\dagger}
\left[ \vec{\sigma} + 4\sin^2 \theta_W \frac{1}{m_e}
(- i\vec{\nabla} - \vec{\sigma} \times \vec{\nabla}) \right]  e
\right)
\right]
\,.
\label{weak nc2}
\end{eqnarray}
This is rearranged to
\begin{eqnarray}
&&
{\cal H}_W = 
\frac{G_F}{\sqrt{2}} \sum_{ij} j_{ij}^{\alpha}j_{ij\,,\alpha}^e
\,, \hspace{0.5cm}
j_{ij}^{\alpha}= \nu_i^{\dagger} \sigma^{\alpha}\nu_j
\,,
\\ &&
\hspace*{-1cm}
j_{ij\,,0}^e = 4 e^{\dagger}
\left(
U_{ei}^*U_{ej} - 
\frac{\delta_{ij}}{2} (1 - 4\sin^2 \theta_W ) + 2i\delta_{ij}
\frac{ \sin^2 \theta_W}{m_e}\vec{\sigma}\cdot\vec{\nabla}
\right)e
\,,
\\ &&
\hspace*{-1cm}
j_{ij\,,k}^e = 4e^{\dagger} \left( \sigma_k
(U_{ei}^*U_{ej} - \frac{1}{2}\delta_{ij})
- 2\delta_{ij}\sin^2 \theta_W\frac{(-i \vec{\nabla} - \vec{\sigma} 
\times \vec{\nabla})_k}{m_e}
\right)e
\,,
\label{weak 4-fermi}
\end{eqnarray}
with $\sigma^{\alpha} = (1\,, \vec{\sigma})$.

\vspace{0.5cm}
\subsection{Dirac neutrino and comparison with Majorana neutrino}

\hspace*{0.5cm}
The relation to the familiar
4-component Dirac equation is explained as follows.
Using a representation of the Clifford algebra that diagonalizes
$\gamma_5$ (its explicit form is given in Appendix A),
the Dirac equation is decomposed into two equations for
two independent 2-spinors, $\varphi$ and $\chi$;
\begin{eqnarray}
&&
(i\partial_t - i\vec{\sigma}\cdot\vec{\nabla}) \varphi = m \chi
\,, \hspace{0.5cm}
(i\partial_t + i\vec{\sigma}\cdot\vec{\nabla}) \chi = m \varphi
\,.
\label{dirac eq}
\end{eqnarray}
Thus, the identification by $\chi = i\sigma_2 \varphi^*$
in the Dirac equation gives the Majorana equation,
eq.(\ref{majorana equation}).

Physical content of these two equations appears different;
only 2 helicity states exist for the Majorana field, a particle being
identical to its anti-particle, unlike distinguishable particle and
anti-particle for the Dirac case.
Whether the neutrino as observed in the $V-A$ weak interaction belongs to
the Majorana case or the Dirac case is the unsettled question facing
fundamental physics.

It is important to theoretically compare the chirality-projected
Dirac field $\psi_D = (1 - \gamma_5)\psi/2$ to the Majorana field.
The relevant 2-component operator corresponding to the momentum eigenstate
$\propto e^{-iE_p t + i \vec{p}\cdot\vec{x}}$ is
\begin{eqnarray}
&&
\hspace*{-1cm}
\psi_{D} = b(\vec{p}\,, h)e^{-i p\cdot x}  u(\vec{p}\,, h) 
+ d^{\dagger}(\vec{p}\,, -h) e^{ip\cdot x} 
\sqrt{\frac{E_p + hp}{E_p - hp}}(-i\sigma_2)u^*(\vec{p}\,, h)
\,.
\label{dirac 1}
\end{eqnarray}
Here $u(\vec{p}\,, h)$ is given by eq.(\ref{m-spinor}).
Anti-particle creation operator $d^{\dagger}(\vec{p}\,, h)$
appears here, which is distinct from the particle creation 
 $b^{\dagger}(\vec{p}\,, h)$.

Comparison of the Majorana solution, eq.(\ref{anti-particle creation}), 
(\ref{m-spinor}) 
and the projected Dirac solution, eq.(\ref{dirac 1}) demonstrates 
the equivalence of the two wave functions,
the difference being the distinction
of the Majorana particle $c^{\dagger}$ and the Dirac 
anti-particle $d^{\dagger}$ creation.
Their distinction appears only via the identical particle effect of two
Majorana fermions, 
an extra term containing $c_2^{\dagger}c_1^{\dagger} = - c_1^{\dagger}c_2^{\dagger}$.
The unique process to
distinguish the massive Majorana from the massive Dirac neutrino is thus
the pair emission $\nu \bar{\nu}$, in which
the anti-symmetrized wave function appears only for
Majorana neutrinos.
We shall later show that this distinction too disappears
in the high energy limit of $E_p \gg m$.
On the other hand,
neither a single nor a pair $\nu\, \nu$ (not $\nu\, \bar{\nu}$) emission 
can tell their distinction even for non-relativistic massive neutrinos,
although they may be able to determine the absolute mass of neutrinos.
In this sense the double beta decay of two accompanying neutrinos 
$\nu_e \nu_e$
is useless for distinction of the Majorana and the Dirac neutrino.
There are thus only two experimental ways to verify the Majorana nature of
neutrinos; the other indirect method is 
to verify the lepton number non-conservation
such as in the neutrinoless double beta decay.

\vspace{0.5cm}
\subsection{Pair emission}

\hspace*{0.5cm}
The idea of using the decay of unstable elementary
particles to verify the Majorana nature of neutrinos 
via identical particle effects is not new;
for instance, see \cite{kpi} and \cite{kotani}.
The problem of this approach is a huge disparity of
energy scales; in both cases of the rare K-meson decay
$K \rightarrow \pi + \nu_{i}\nu_{j}$ \cite{kpi}
and the muon decay $\mu^+ \rightarrow e^+ \nu_e \bar{\nu}_{\mu}$ 
\cite{kotani},
the mass difference is much larger than anticipated 
neutrino masses, and even if events of this process are accumulated
statistically, there is no sensible way to precisely determine
the neutrino mass.

On the other hand, the atomic energy difference is closer to
the neutrino mass scale;
\begin{eqnarray}
&&
\Delta E_{n_1\,, n_2} \sim 13.6 eV \times (\frac{1}{n_2^2} - \frac{1}{n_1^2})
\end{eqnarray}
which reduces to 
$27\, meV \Delta n/ (n/10)^3 $ for $|n_1 - n_2| \sim n \gg 1$
for Rydberg states.

Let us first discuss the neutrino pair $\nu_2 \bar{\nu}_1$
emission of Dirac particles.
Note that the anti-particle notation $\bar{\nu}_1$ 
is necessary only for the Dirac case. 
The Dirac pair  emission 
$\propto b^{\dagger}(\vec{p}_2 \,, h_2)d^{\dagger}(\vec{p}_1 \,, h_1)$ 
is governed by $e^{i(\vec{p}_1 + \vec{p}_2)\cdot \vec{x}}$ times 
the current matrix element,
\begin{eqnarray}
&&
\hspace*{1cm}
j_D^{\alpha}(\vec{p}_1 h_1\,, \vec{p}_2 h_2) 
= - \sqrt{\frac{E_1 + h_1p_1}{E_1 - h_1p_1}} 
u^{\dagger}(\vec{p}_2 \,, - h_2)\sigma^{\alpha}i\sigma_2
u^*(\vec{p}_1 \,, h_1)
\,.
\end{eqnarray}
The pair emission rate contains a neutrino current product;
\begin{eqnarray}
&&
\hspace*{-1cm}
j_D^{\alpha}(j_D^{\beta})^{\dagger}
= \frac{1}{16} (1 + h_2 \frac{p_2}{E_2})(1 + h_1 \frac{p_1}{E_1})
{\rm tr}\;
(1 - h_2 \frac{\vec{\sigma}\cdot\vec{p}_2}{p_2}) \sigma^{\alpha}
(1 - h_1 \frac{\vec{\sigma}\cdot\vec{p}_1}{p_1}) \tilde{\sigma}^{\beta}
\,,
\nonumber \\ &&
\end{eqnarray}
with $\tilde{\sigma}^{\beta} = (1\,, - \vec{\sigma})$.

The neutrino pair current given above is to be multiplied 
by the electron current product.
After this multiplication, the helicity summed quantity for the Dirac case is
\begin{eqnarray}
&&
\sum_{h_1 h_2} |j_D \cdot j^e|^2 = 
\frac{1}{2}\left(
(1 + \frac{\vec{p}_1\cdot\vec{p}_2}{E_1 E_2})
j^e_{0}(j^e_{0})^{\dagger}
+ (1 - \frac{\vec{p}_1\cdot\vec{p}_2}{E_1 E_2})
\vec{j}^e (\vec{j}^e)^{\dagger}
\right.
\nonumber \\ &&
+ 2 \Re \frac{\vec{p}_1 \cdot \vec{j}^e\vec{p}_2 \cdot (\vec{j}^e)
^{\dagger}}{E_1 E_2}
- 2(\frac{\vec{p}_1}{E_1} + \frac{\vec{p}_2}{E_2}) \cdot 
\Re j_0^e(\vec{j}^e)^{\dagger} 
\nonumber \\ &&
\left.
+ 2 \frac{\vec{p}_1\times \vec{p}_2}{E_1 E_2} \cdot 
\Im j_0^e(\vec{j}^e)^{\dagger}
+ 2(\frac{\vec{p}_1}{E_1} - \frac{\vec{p}_2}{E_2})\cdot 
\Re \vec{j}^e \times \Im \vec{j}^e
\right)
\,.
\end{eqnarray}
The last two quantities in proportion to imaginary parts 
of the current product may contain  CP-odd effects.

Let us next discuss the Majorana pair emission.
The Majorana pair emission operator,
$c^{\dagger}(\vec{p}_2\,, h_2)c^{\dagger}(\vec{p}_1\,, h_1)$,
gives a matrix element of two anti-symmetrized wave functions
due to the anti-commutation of the Majorana field.
The neutrino current for the pair emission is thus
$e^{i(\vec{p}_1 + \vec{p}_2)\cdot \vec{x}}$ times 
\begin{eqnarray}
&&
j_M^{\alpha} (\vec{p}_1 h_1\,, \vec{p}_2 h_2) = 
-i \sqrt{\frac{E_2 - h_2p_2}{E_2 + h_2p_2}}
u^{\dagger}(\vec{p}_1 \,, h_1)\sigma^{\alpha}\sigma_2
u^*(\vec{p}_2 \,, - h_2)
\nonumber \\ &&
+ i\sqrt{\frac{E_1 + h_1p_1}{E_1 - h_1p_1}}
u^{\dagger}(\vec{p}_2 \,, - h_2)\sigma^{\alpha}\sigma_2
u^*(\vec{p}_1 \,, h_1)
\,.
\label{anti-symmetrized wf}
\end{eqnarray}

To derive the rate, one multiplies the electron current $j_e^{\alpha}$
and takes its square.
After a little algebra, one finds for the relevant quantity
of the neutrino part,
\begin{eqnarray}
&&
\hspace*{1cm}
j_M^{\alpha} (j_M^{\beta})^{\dagger} 
= \frac{1}{16}(1 + h_1 \frac{p_1}{E_1})(1 + h_2 \frac{p_2}{E_2}) \times
\nonumber \\ &&
{\rm tr}\;
(1 - h_2 \frac{\vec{\sigma}\cdot\vec{p}_2}{p_2}) \sigma^{\alpha}
(1 - h_1 \frac{\vec{\sigma}\cdot\vec{p}_1}{p_1}) \tilde{\sigma}^{\beta}
+ (1 \leftrightarrow 2)
\\ &&
+ \frac{m_1 m_2}{16 E_1 E_2}
{\rm tr}\;
(1 - h_2 \frac{\vec{\sigma}\cdot\vec{p}_2}{p_2}) \sigma^{\alpha}
(1 - h_1 \frac{\vec{\sigma}\cdot\vec{p}_1}{p_1}) \sigma^{\beta}
+ (1 \leftrightarrow 2)
\,.
\label{interference}
\end{eqnarray}
The last two terms of eq.(\ref{interference}) are the interference term 
proper to identical fermions.

It is evident that without the interference terms $\propto m_1 m_2 /(E_1 E_2)$
in (\ref{interference}), 
the Dirac and the Majorana
emission rates are identical, by considering an extra factor 1/2 
for the Majorana case,
which is necessary after the phase space integration 
because the same configuration
of identical particles are counted twice.
Hereafter we devide the Majorana contribution
by 2, anticipating this overcouting beforehand.

The  helicity summed interference term thus becomes
\begin{eqnarray}
&&
 \frac{m_1 m_2}{2 E_1 E_2} \left( j_0^e (j_0^e)^{\dagger} 
- \vec{j}^e\cdot (\vec{j}^e)^{\dagger}\right)
\,.
\label{helicity summed interference}
\end{eqnarray}
Hence the current product in the rate for the Majorana emission
is 
\begin{eqnarray}
&&
\hspace*{-1cm}
\sum_{h_1 h_2} |j_M\cdot j^e|^2 
= \sum_{h_1 h_2} |j_D \cdot j^e|^2
+ \frac{m_1 m_2}{2 E_1 E_2} \left( j_0^e (j_0^e)^{\dagger} 
- \vec{j}^e\cdot (\vec{j}^e)^{\dagger}\right)
\,.
\label{Majorana rate 0}
\end{eqnarray}
The interference term is CP-even.
Hence the CP violating effect is identical in the Dirac and the Majorana
cases, for the neutrino pair emission.

As an illustration, let us work out the current product by
neglecting $1/m_e$ terms and taking $\sin^2 \theta_W = 1/4$.
We call this the leading approximation
of $1/m_e$ expansion.
The spin and the orbital part of the electron current is
separated as
\begin{eqnarray}
&&
j_0^e = \delta_{ss'} a_{if}(\vec{p}_1 + \vec{p}_2)c_{ij}^{(0)}
\,, \hspace{0.5cm}
\vec{j}^e = \langle s' | \vec{\sigma} | s \rangle
a_{if}(\vec{p}_1 + \vec{p}_2)c_{ij}^{(s)}
\,,
\\ &&
a_{if}(\vec{p}_1 + \vec{p}_2)
= \langle f | e^{-i (\vec{p}_1 + \vec{p}_2) \cdot\vec{x}}| i \rangle
\,,
\\ &&
c_{ij}^{(0)} = U_{ei}^*U_{ej}
\,, \hspace{0.5cm}
c_{ij}^{(s)} = U_{ei}^*U_{ej} - \frac{1}{2}\delta_{ij}
\,.
\end{eqnarray}
Furthermore, we consier the spin averaged rate such that
\begin{eqnarray}
&&
\hspace*{-1cm}
\frac{1}{2}\sum_{s s'} j^e_k (j^e_l)^{\dagger}
= \delta_{kl}
|a_{if}(\vec{p}_1 + \vec{p}_2)|^2 |c_{ij}^{(s)}|^2
\,,
\hspace{0.5cm}
\frac{1}{2}\sum_{s s'} j^e_0 (j^e_0)^{\dagger}
= |a_{if}(\vec{p}_1 + \vec{p}_2)|^2 |c_{ij}^{(0)}|^2
\,. \nonumber
\end{eqnarray}

The result of the spin average for the Dirac and the Majorana
cases is
\begin{eqnarray}
&&
\hspace*{-1cm}
\frac{1}{2}\sum_{s s'}\sum_{h_1 h_2} |j_D\cdot j^e|^2 = 
|a_{if}(\vec{p}_1 + \vec{p}_2)|^2
\frac{1}{2}\left(
(1 + \frac{\vec{p}_1\cdot\vec{p}_2}{E_1 E_2}) |c_{ij}^{(0)}|^2
+ (3 - \frac{\vec{p}_1\cdot\vec{p}_2}{E_1 E_2})|c_{ij}^{(s)}|^2
\right.
\nonumber \\ &&
\hspace*{1cm}
\left.
+ 
2(\frac{\vec{p}_1}{E_1} - \frac{\vec{p}_2}{E_2})\cdot 
{\mathrm{Re}} \vec{j}^e \times \Im \vec{j}^e |c_{ij}^{(s)}|^2
\right)
\,,
\\ &&
\frac{1}{2}\sum_{s s'}\sum_{h_1 h_2} |j_M\cdot j^e|^2 - 
\frac{1}{2}\sum_{s s'}\sum_{h_1 h_2} |j_D\cdot j^e|^2
\nonumber \\ &&
\hspace*{1cm}
= 
- |a_{if}(\vec{p}_1 + \vec{p}_2)|^2
\frac{m_1 m_2}{2E_1 E_2}(3 |c_{ij}^{(s)}|^2
- |c_{ij}^{(0)}|^2)
\,.
\label{majorana h-sum}
\end{eqnarray}

The long wavelength approximation for the neutrino makes this
formula much simpler, allowing the replacement,
\begin{eqnarray}
&&
|a_{if}(\vec{p}_1 + \vec{p}_2)|^2 =
|\langle f | e^{-i (\vec{p}_1 + \vec{p}_2) \cdot\vec{x}}| i \rangle|^2
\rightarrow 1
\,.
\end{eqnarray}
This is valid if the wavelength of the neutrino
$\lambda_{\nu} \gg $ atomic size, or
$p_{\nu} \ll Z \alpha m_e/n^2$.
Since our main interest is in the region of the neutrino mass,
this means, with $m_{\nu} \leq p_{\nu}$,
\begin{eqnarray}
&&
m_{\nu} \ll \frac{Z \alpha m_e}{n} \sim 3.7 keV\, \frac{Z}{n^2}
\,.
\end{eqnarray}
This relation holds in the following discussion for
laser irradiated process, but may not for pair emission
from Rydberg atoms.
The condition for the long wavelength approximation is significantly
modified for Rydberg atoms.

\vspace{1cm}
\section{Kinematics of pair decay}

\vspace{0.5cm}
\subsection{What can be measured from the pair decay}

\hspace*{0.5cm}
A merit of the process of neutrino pair emission from
excited atoms is a great variety of measurable quantities related to
the neutrino mass parameter.
The neutrino pair $\nu_i \nu_j$ can be any combination $ij$ 
of mass eigenstates,
as is clear from coexistence of the charged and the neutral current
interaction.
From the energy threshold position one can determine a combination of
the neutrino mass $m_i + m_j$, while the strength of the rate
gives the mixing parameter in the form,
$|U_{e i}^*U_{e j}|$.
Since any pair $ij$ (altogether 6 channels) is conceivable, 
there is a great many combinations.
This is why we phrased our experimental approach as the Mneutrino spectroscopy.
The situation is quite different from the neutrinoless double
beta decay in which one concentrates only on a combination
of parameters $|\sum_i U_{e i}U_{e i}m_i^2|$, which however
attempts to discover the important issue of
lepton number violation.

Ideally, one can determine all masses $m_i$ with a bonus of
experimental redundancy.
In particular, we would like to emaphasize that this is the first
opportunity to probe the smallest mass $m_1$.
Furthermore, we may explore a possibly very small mixing
factor $|U_{e 3}|$, which indicates how much the heaviest neutrino
is mixed in the flavor $\nu_e$.
Both are important since other experimental methods may have
no good handle on these quantities.

\vspace{0.5cm}
\subsection{Phase space factor}

\hspace*{0.5cm}
When the pair emission occurs between 2 levels of energy difference
$\Delta$, the rate is given by
the phase space factor of 2 massive neutrinos $\nu_i \nu_j$.
The rate is
\begin{eqnarray}
&&
\hspace*{-1cm}
\Gamma_{ij}^{M\,, D}(\Delta) = 8G_F^2
\int \frac{d^3 q_1 d^3 q_2}{(2\pi)^5}
\delta (E_1 + E_2 - \Delta ) 
\frac{1}{2}\sum_{s s'}\sum_{h_1 h_2} |c_{ij}j_{M\,,D}\cdot j^e|^2 
\,.
\label{rate formula 1}
\end{eqnarray}
The constant $c_{ij}$'s are different, depending on whether
the electron transition involves spin-flip (F) or no flip (NF).
They are in the leading $1/m_e$ approximation,
\begin{eqnarray}
&&
c_{ij} = c_{ij}^{(s)} = U_{e_i}^*U_{e_j} - \frac{\delta_{ij}}{2} \; ({\rm F})
\,, \hspace{0.5cm}
c_{ij} = c_{ij}^{(0)} = U_{e_i}^*U_{e_j} \; ({\rm NF})
\,,
\\ &&
\sum_{ij}|c_{ij}^{(s)}|^2 = \frac{3}{4}
\,, \hspace{0.5cm}
\sum_{ij}|c_{ij}^{(0)}|^2 = 1
\,.
\end{eqnarray}
The kinematical factor is defined, when the matrix element is ignored, as
\begin{eqnarray}
&&
f_{ij}^{(1)}(\Delta) = 
\int \frac{d^3 q_1 d^3 q_2}{(2\pi)^5}
\delta(E_1 + E_2 - \Delta)
\nonumber \\ &&
=
\frac{1}{2\pi^3}
\int_{m_i}^{\Delta - m_j}dE_1 \,E_1 (\Delta - E_1) [(E_1^2 - m_i^2)
\left( (\Delta - E_1)^2 - m_j^2 \right)]^{1/2}
\,.
\label{phase space for pair emission}
\end{eqnarray}
Near the threshold $\Delta \approx  m_i + m_j$
\begin{eqnarray}
&&
f_{ij}^{(1)}(\Delta) 
\approx \frac{1}{8\pi^2}(m_i m_j)^{3/2} (\Delta - m_i - m_j)^2
\,.
\end{eqnarray}
Note that the distinction of Majorana and Dirac particles
appears only when $i = j$, since with $i \neq j$ two
neutrinos have different masses and are not identical
particle.

For illustration we shall give the rate near
the threshold in the leading approximation of $1/m_e$,
further suppressing the correlation of electron transition
amplitude with neutrino momenta,
\begin{eqnarray}
&&
\Gamma_{ij}^{M}(\Delta) \approx
\frac{G_F^2}{8\pi^2}|c_{ij}^{(0)}|^2
(m_i m_j)^{3/2} (\Delta - m_i - m_j)^2
\,,
\\ &&
\Gamma_{ij}^{D}(\Delta) \approx
\frac{G_F^2}{16\pi^2}(|c_{ij}^{(0)}|^2 + 3|c_{ij}^{(s)}|^2)
(m_i m_j)^{3/2} (\Delta - m_i - m_j)^2
\,.
\label{pair decay rate}
\end{eqnarray}

\vspace{1cm}
\section{Laser irradiated pair decay}

\vspace{0.5cm}
\hspace*{0.5cm}
The weak rate scales with the available energy as $E^5$.
This means that for a small available energy the rate is very small.
For the case of the neutrino pair emission from excited atomas,
the phase space factor is
\begin{eqnarray}
&&
\frac{G_F^2 E^5}{15 \pi^3} \sim 3.3 \times 10^{-34} s^{-1}\,
(\frac{E}{eV})^5
\,,
\label{rough rate}
\end{eqnarray}
which should be multiplied by the matrix element squared.
One clearly needs some enhancement mechanism even to hope
for detectablity of the neutrino pair emission.
We shall discuss in this section laser irradiated pair emission
using a resonance effect.

Let us first estimate very crudely how much enhancement may
be expected for resonant processes.
Suppose that the neutrino pair emission from
a metastable atom of lifetime $1/\gamma$ is triggered by
laser irradiation of flux $F_0$.
The rate for photon absorption is $\sigma F_0$ with
$\sigma$ the photo-absorption cross section, and
this irradiation is effective for the duration of lifetime.
Putting these factors together, one might naively
expect laser irradiated
rate of order, $\sigma F_0 \Gamma_{\nu \bar{\nu}}/\gamma$,
with $\Gamma_{\nu \bar{\nu}}$ the rate of order (\ref{rough rate}).

Another important factor is the energy resolution $\Delta \omega$
of laser.
Convolution of the laser spectral function with
the Breit-Wigner resonance function of the natural
width $\gamma$ leads to a factor of order,
\begin{eqnarray}
&&
\frac{\omega_0^2}{\gamma \Delta \omega}
= 1.5 \times 10^{24}
\frac{10^{-9}\omega_0}{\Delta \omega}\frac{\omega_0}{1 eV}
\frac{1 s^{-1}}{\gamma}
\,,
\end{eqnarray}
with $\omega_0$ the laser central frequency.
It was assumed that $\Delta \omega \gg \gamma$, which is usually valid.
We shall take a standard laser flux of order,
\begin{eqnarray}
&&
F_0 = \frac{W mm^{-2}}{eV} \approx 6.2 \times 10^{20} cm^{-2} sec^{-1}
\approx 1.6 \times 10^{-4} (eV)^3
\,,
\label{typical laser flux}
\end{eqnarray}
and the photo-absorption cross section of order, $nm^2$.
The laser beam power $P$ is related to the number flux $F_0$ by
$P = \omega_0 F_0$, with $\omega_0 = \hbar \,\times$ the laser
frequency.

This leads to a typical laser irradiated rate of order,
\begin{eqnarray}
&&
\hspace*{-1cm}
\frac{G_F^2 E^5}{15 \pi^3}\frac{\omega_0\sigma F_0}{\Delta \omega \gamma}
\sim 2 \times 10^{-18} s^{-1} \frac{1 s^{-1}}{\gamma}
\frac{\sigma}{nm^2}(\frac{E}{1 eV})^5
(10^{-9}\frac{\omega_0}{\Delta \omega})\frac{P}{W mm^{-2}}
\,.
\label{naive laser rate}
\end{eqnarray}
In the following we shall give a more concrete estimate
for a particular process of laser irradiated pair emission.

We consider \cite{fnns} laser irradiated neutrino pair emission 
from metastable ions or atoms $| I^{*} \rangle$,
\begin{eqnarray}
&&
\gamma + I^{*} \rightarrow I^{**} +  \nu_i \nu_j
\,,
\end{eqnarray}
where the final state $|I^{**}\rangle$ is a short-lived excited state,
as dipicted in Figure 1 \cite{fukumi}.
Detection of $|I^{**}\rangle$, presumably via radiative decay
into the ground state, gives a signature of this weak process.
A measurement of parity violating quantity is highly desirable
for the background rejection.
\begin{figure}[htbp]
  \centerline{\includegraphics{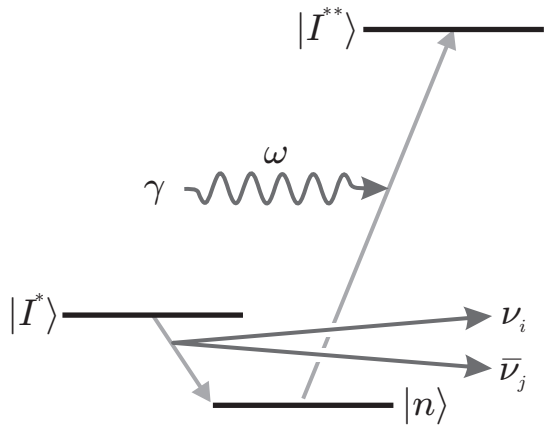}}
  \caption{Atomic level structure and laser irradiated neutrino pair emission}
  \label{Fig 1}
\end{figure}

The intermediate state $|I^n \rangle$ is chosen to lie energetically below
the initial state $|I^{*} \rangle$ by an amount
of the paired neutrino energy $E_i + E_j$ with $ E_i = \sqrt{\vec{p}_i^2 + m_i^2}$.
The laser frequency is tuned to the next step of
radiative transition from $|I^n \rangle$ 
to the final state $| I^{**} \rangle$.
The Breit-Wigner resonance factor
\begin{eqnarray}
&&
\frac{1}{(E_* - E_n - E_i - E_j)^2 + \gamma^2/4}
= \frac{1}{(E_{**} - E_n - \omega)^2 + \gamma^2/4}
\,,
\end{eqnarray}
with the energy conservation $E_* + \omega = E_{**}+ E_i + E_j$,
gives a large enhancement at the threshold $E_i + E_j = m_i + m_j$,
if
\begin{eqnarray}
&&
E_n \approx E_* - m_i - m_j
\,, \hspace{0.5cm}
\omega \approx E_{**} - E_* + m_i + m_j
\,.
\end{eqnarray}
Thus the laser initiated pair emisssion has the threshold at
the laser energy of
\begin{eqnarray}
&&
\omega_{th} = E_{**} - E_* + m_i + m_j
\,.
\end{eqnarray}
The threshold location $\omega_{th}$ is expected to be measured with a good
precision of the laser frequency.

The initial state $|I^* \rangle $ must be a metastable excited state, 
for instance $O[2m_3]$ above another state $|I^n\rangle$.
The intermediate state $|I^n\rangle$ can be either a ground state, or better,
another metastable state.
The width factor given by
$\gamma^2 = \gamma_*^2 + \gamma_n^2$ is a sum of the initial and 
the intermediate state contributions.
We assume that both of these widths are of order $1 sec^{-1}$ or smaller.
There are many candidate atoms or ions of this kind.
Another important assumption is that we prepare depletion
of the intermediate state $|I^n\rangle$, 
since the laser excitation of this state
to $|I^{**} \rangle$ is the crucial signature of the pair emission process.

To lowest order of the weak interaction, the rate for
the pair emission $\nu_i \nu_j$ of mass eigenstates
when a laser of flux $F(\omega)$ is irradiated,
is given by
\begin{eqnarray}
&&
\hspace*{-1cm}
\Gamma^{M\,,D}_{ij}(\omega) = F(\omega) 
\frac{8G_F^2 e^2 \omega |\langle I^{**}|\vec{x}| I^n \rangle|^2}
{3[(\omega - \Delta_{fn})^2 + \gamma^2/4]}
f^{M\,,D}_{ij}(\omega - \Delta_{fi})
\,,
\label{rate formula 1}
\\ &&
2 f^D_{ij}(\Delta)= 
f_{ij}^{(1)}(\Delta) (|c_{ij}^{(0)}|^2 + 3|c_{ij}^{(s)}|^2 )
\,,
\\ &&
2 f^M_{ij}(\Delta) - 2 f^D_{ij}(\Delta)
= f_{ij}^{(2)}(\Delta) (|c_{ij}^{(0)}|^2 -3|c_{ij}^{(s)}|^2 )
\,,
\\ &&
f_{ij}^{(2)}(\Delta)
= \int \frac{d^3 q_1 d^3 q_2}{(2\pi)^5}\delta(E_1 + E_2 - \Delta) 
\frac{m_i m_j}{E_1 E_2}
\,,
\end{eqnarray}
where $f_{ij}^{(1)}$ is defined by (\ref{phase space for pair emission}),
and 
\begin{eqnarray}
&&
\Delta_{fn} = E_{**} - E_n
\,, \hspace{0.5cm}
\Delta_{fi} = E_{**} - E_*
\,.
\end{eqnarray}
The Majorana and the Dirac difference 
\(\:
\propto f_{ij}^{(2)}(\omega - \Delta_{fi})
\times (|c_{ij}^{(0)}|^2 -3|c_{ij}^{(s)}|^2)
\,.
\:\)

When integrated over a laser spectral function of
the energy resolution $\Delta \omega \gg \gamma$,
\begin{eqnarray}
&&
\int d\omega \frac{F(\omega)}{(\omega - \Delta)^2 + \gamma^2/4}
\sim \frac{2\pi F_0}{\gamma \Delta \omega}
\,,
\end{eqnarray}
the rate is 
\begin{eqnarray}
&&
\Gamma^M_{ij}(\omega) = \frac{64\pi^2 \alpha G_F^2}{3}
\frac{\omega |\langle \vec{x} \rangle |^2 F_0}{\gamma \Delta \omega}
f^M_{ij}(\omega - \Delta_{fi})
\,.
\end{eqnarray}
The dipole strength is related to the natural width,
\begin{eqnarray}
&&
\gamma_r = \frac{4\alpha}{3} \omega^3 |\langle \vec{x} \rangle |^2 
\,,
\end{eqnarray}
which gives
\begin{eqnarray}
&&
\Gamma^M_{ij}(\omega_0) = \frac{16\pi^2 G_F^2 \gamma_r F_0}
{\omega_0^2 \gamma \Delta \omega}
f^M_{ij}(\omega_0 - \Delta_{fi})
\nonumber \\ &&
= 16\pi^2 G_F^2 \frac{F_0}{\omega_0^2 \Delta \omega}\frac{\gamma_r}{\gamma}
f^M_{ij}(\omega_0 - \Delta_{fi})
\,.
\end{eqnarray}
We assume that the laser tuning is complete.
In this last formula three factors are separated;
the laser quality factor $F_0/(\omega_0^2 \Delta \omega)$, 
the atomic factor $\gamma_r /\gamma$, and
the neutrino kinematical factor $f^M_{ij}$.

Near the threshold,
denoting the tuned frequency $\omega_0$ by $\omega$,
\begin{eqnarray}
&&
\hspace*{-1cm}
f_{ij}^M(\omega - \Delta_{fi}) =
\frac{1}{8\pi^2}|c_{ij}^{(0)}|^2
(m_i m_j)^{3/2} (\omega - \Delta_{fi} - m_i - m_j)^2
\,,
\\ &&
\hspace*{-1cm}
f_{ij}^D(\omega - \Delta_{fi}) =
\frac{1}{16\pi^2}(3|c_{ij}^{(s)}|^2 + |c_{ij}^{(0)}|^2)
(m_i m_j)^{3/2} (\omega - \Delta_{fi} - m_i - m_j)^2
\,.
\end{eqnarray}

To compute a reference rate let us take $13/4$ for the 
asymptotic value of the factors of
$|c_{ij}|^2$, which gives a basic unit of the rate,
\begin{eqnarray}
&&
\hspace*{-1cm}
\frac{13}{64 \pi^2}
\frac{G_F^2 F_0 \gamma_r }
{\omega^2 \Delta \omega \gamma} (eV)^5 = 5.0 \times 10^{-19} s^{-1}
\frac{P}{W mm^{-2}}(\frac{eV}{\omega})^4
\frac{10^{-9}\omega}{\Delta \omega}\frac{10^{-9}\gamma_r}{\gamma}
\,.
\end{eqnarray}
We have in mind radiative rates of order,
$1/\gamma_r \sim 1 ns $ and $1/\gamma \sim  1s$.
Taking the energy scale at $0.3 eV$, about 3 times
the pair mass of the heaviest neutrino $50 meV$,
then gives the rate of order,
\begin{eqnarray}
&&
1 \times 10^{-21} s^{-1}
\frac{f_{ij}^M(\omega)}{(0.3 eV)^5}
\frac{P}{W mm^{-2}}(\frac{eV}{\omega})^4
\frac{10^{-9}\omega}{\Delta \omega}\frac{10^{-9}\gamma_r}{\gamma}
\,.
\end{eqnarray}
This corresponds to $\approx 1 $ event/day for $10^{16}$ target atoms.
With a more experimental effort of improvement such
as the use of the resonator for laser irradiation,
an enhancement of $\approx 10^3- 10^4$ may be expected and would much help
for the improved event rate. 

The threshold suppression is large due to
the square factor $(\omega - \Delta_{fi} - m_i - m_j)^2$, 
but the rate rises towards
\begin{eqnarray}
&&
\frac{2 G_F^2 \omega^5}{15\pi}
\left(3 |c_{ij}^{(s)}|^2 + |c_{ij}^{(0)}|^2 \right)
\frac{\gamma_r F_0}{\gamma \omega^2 \Delta \omega}
\,.
\end{eqnarray}
This pattern repeats for each pair $i\, j$,
and finally approaches at much larger $\omega \gg 2m_3$,
\begin{eqnarray}
&&
\frac{2 G_F^2 \omega^5}{15\pi}
(\frac{9}{4} + 1)
\frac{\gamma_r F_0}{\gamma \omega^2 \Delta \omega}
\,,
\end{eqnarray}
where $9/4$ comes from the spin flip term, while $1$ from
the non-flip term.
To determine neutrino massses, it is necessary to
fit experimental data of the threshold rise up to an intermediate
energy range. It is then important to
have a large statistics data with a reasonable precision
in the vicinity of the threshold,
typically away from the threshold a few to several times the sum
$m_i + m_j$.

The neutrino mass spectroscopy may proceed step by step.
First, the laser frequency dependence, for instance 
$\propto (\omega - \Delta_{fi} - m_i - m_j)^2$ near the threshold,
may be used to determine mass parameters $m_i$.
Simultaneous with or even prior to
$m_3$ determination at the threshold $\nu_3 \,\nu_3$,
distinction of the Majorana and the Dirac cases is presumably possible
at $\omega - \Delta_{fi} \sim O[6 m_3] \approx 0.3 eV$.
Once the mass determination is done, one proceeds to determine mixing angles 
by
measurement of the absolute rate.
For instance, the sensitivity to the smallest, unknown angle $\theta_{13}$
is large at the threshold of $\omega = \Delta_{fi} + m_3 + m_1$,
since the relevant factor has a large coefficient;
\begin{eqnarray}
&&
3|c_{13}^{(s)}|^2 + |c_{13}^{(0)}|^2 \sim 2.9 \sin^2 \theta_{13}
\,.
\end{eqnarray}
For the heierarchical mass pattern, $m_3 + m_1 \sim 50 meV$.
The rate at this threshold is amaller by a factor $\sim 1/32$ than
at the the threshold $2m_3 \sim 100 meV$.
Although smaller in the rate, the $\theta_{13}$ 
measurement may be possible.

We conclude that for precision determination of
absolute values of $m_i \,, (i= 1\,, 2\,, 3)$ 
and $\theta_{13}$, pair emissions of $(\nu_3 \nu_3) \,, 
(\nu_3 \nu_2) \,, (\nu_3 \nu_1)$ near their thresholds
are channels we recommend.

A quantitative Majorana-Dirac distinction may be much
helped by noting the rate difference at each threshold $ij$;
\begin{eqnarray}
&&
\Gamma^M_{ij}(\omega_0) - \Gamma^D_{ij}(\omega_0)
= 8\pi^2 G_F^2 \frac{\gamma_r F_0}{\gamma \omega_0^2 \Delta \omega}
(|c_{ij}^{(0)}|^2 - 3 |c_{ij}^{(s)}|^2)
\,,
\\ &&
|c_{ij}^{(0)}|^2 - 3 |c_{ij}^{(s)}|^2 =
- 2|U_{ei}|^2|U_{ej}|^2  \,,\; {\rm for}\; i \neq j
\\ &&
\hspace*{1cm}
= - 2|U_{ei}|^4 + 3|U_{ei}|^2 - \frac{3}{4}
\,,\; {\rm for}\; i = j
\,.
\end{eqnarray}

For a reference, we give a complete rate formula including
all pair channels;
\begin{eqnarray}
&&
\Gamma^M(\omega_0) =
\frac{2G_F^2}{\pi^2}\frac{\gamma_r}{\omega_0^2} \int d\omega \sum_{ij}
\theta (\omega - \Delta_{fi} - m_i - m_j ) 
\nonumber \\ &&
\hspace*{1cm}
\times
\frac{ F(\omega \,; \omega_0 \,, \Delta \omega)}
{(\omega - \Delta_{fn})^2 + \gamma^2/4}
\nonumber \\ &&
\times
\left[
(|c_{ij}^{(0)}|^2 + 3|c_{ij}^{(s)}|^2 )
(\omega - \Delta_{fi})^5
Y_{ij}\left( \frac{m_i}{\omega - \Delta_{fi}}\,, 
\frac{m_j}{\omega - \Delta_{fi}}\right)
\right.
\nonumber \\ &&
\left.
+ \delta_{ij}(|c_{ij}^{(0)}|^2 - 3|c_{ij}^{(s)}|^2 )
m_i m_j (\omega - \Delta_{fi})^3
Z_{ij}\left( \frac{m_i}{\omega - \Delta_{fi}}\,, 
\frac{m_j}{\omega - \Delta_{fi}}\right)
\right]
\,,
\label{complete rate formula}
\\ &&
Y_{ij}(\epsilon_i\,, \epsilon_j) = 
\int_{\epsilon_i}^{1 - \epsilon_j} dx x (1-x)
\sqrt{(x^2 - \epsilon_i^2) \left( (1 - x)^2 - \epsilon_j^2 \right)}
\,, 
\\ &&
Z_{ij}(\epsilon_i\,, \epsilon_j) = 
\int_{\epsilon_i}^{1 - \epsilon_j} dx 
\sqrt{(x^2 - \epsilon_i^2) \left( (1 - x)^2 - \epsilon_j^2 \right)}
\,.
\end{eqnarray}
We used a notation of laser spectral function 
$F(\omega \,; \omega_0 \,, \Delta \omega)$, which has 
a central frequency $\omega_0$ and an energy resolution $\Delta \omega$.
In the case of the Dirac neutrino the
last term $m_i m_j Z_{ij}$ in (\ref{complete rate formula}) is missing.

\begin{figure}[htbp]
  \centerline{\includegraphics{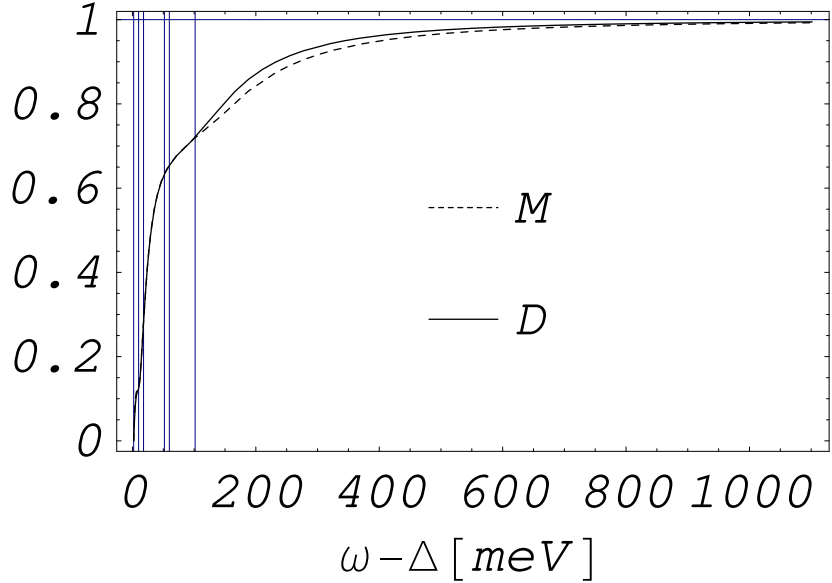}}
  \caption{Ratio of laser irradiated pair emission rate for Majorana (M) and
Dirac(D) cases to the massless rate. Vertical lines indicate 6 threshold locations.}
  \label{Fig 2}
\end{figure}

Under the assumption of $\Delta \omega \gg \gamma$, this rate
becomes simplified as
\begin{eqnarray}
&&
\Gamma^{M}(\omega) =  
\frac{4 G_F^2 F_0}{\pi \omega^2 \Delta \omega}
\frac{\gamma_r}{\gamma} \sum_{ij}
\theta (\omega - \Delta_{fi} - m_i - m_j ) 
\int_{m_i}^{\omega - \Delta_{fi} - m_j} dE_1 I(E_1)
\,,
\\ &&
\hspace*{-1cm}
I(E_1) = 
(|c_{ij}^{(0)}|^2 + 3|c_{ij}^{(s)}|^2 )
E_1(\omega - \Delta_{fi} - E_1)
\sqrt{(E_1^2 - m_i^2)(\,(\omega - \Delta_{fi} - E_1)^2 - m_j^2\,)}
\nonumber \\ &&
\hspace*{1cm}
+\delta_{ij}(|c_{ij}^{(0)}|^2 - 3|c_{ij}^{(s)}|^2 )
m_i m_j \sqrt{(E_1^2 - m_i^2)(\,(\omega - \Delta_{fi} - E_1)^2 - m_j^2\,)}
\,.
\end{eqnarray}
We plot in Figure 2 the ratio of two rates;
the Majorana rate $\Gamma^M(\omega)$ and the Dirac rate $\Gamma^D(\omega)$ 
divided by the rate of massless neutrino pair emission.
The paremeters used for this figure are
\begin{eqnarray*}
&&
\sin^2 \theta_{12} = 0.35 \,,
\hspace{0.3cm}
\sin^2 \theta_{13} = 0.032 \,,
\\ &&
m_1 = 1.0 meV \,, \hspace{0.3cm}
m_2 = 9.0 meV \,, \hspace{0.3cm}
m_3 = 50.8 meV \,,
\end{eqnarray*}
constrained and allowed by neutrino oscillation data.
With this small mass $m_1$ the energy region in which
the Majorana rate is larger than the Dirac rate
is restricted to a small region below $m_1 + m_2$.
When $m_1$ is larger, say $m_1 \geq 2 meV$,
the Majorana dominance persists up to slightly
above the $2m_3$ threshold.
The Majorana dominance is also sensitive to a value of $\sin^2 \theta_{12}$
\cite{oka}.

These rates presented here may be an overestimate for a dense gaseous target, 
since the energy resolution in this case is governed by a usually much
larger collisional width,
\begin{eqnarray}
&&
\gamma_{coll} \approx 1 s^{-1}\frac{P}{10^{-7}Torr}
\sqrt{\frac{T}{300 K}}\frac{\sigma v}{1 nm^2}
\,.
\end{eqnarray}

\vspace{1cm}
\section{Neutrino pair emission from circular Rydberg states}

\vspace{0.5cm} 
\hspace*{0.5cm}
\subsection{Circular Rydberg states}

\hspace*{0.5cm} 
Circular Rydberg states \cite{circular states} 
are highly excited;
in addition to a large principal quantum number $n$, it has
the highest angular momentum, $l = |m| = n -1 $.
These states have the least overlap with the atomic core of charge $(Z - 1)e$.
Its size $\langle r \rangle \approx n^2 a_B$ with
a dispersion $\sqrt{\langle (\Delta r)^2 \rangle} \approx \sqrt{n^3/2} a_B$, 
and the average momentum 
$\langle p \rangle \approx 1/(n a_B)$ with a large dispersion.
Thus, the circular Rydberg state is almost classical, and
the system is approximately described by the hydrogen-like
Coulomb potential of charge $e$.
A great merit of circular Rydberg atoms is that lifetime of radiative
decay is very long, scaling as $n^5$
with the principal quantum number $n$, and $\sim O[1ms]$ for $n \sim 25$
\cite{ry-decay rate}.

The wave function $\psi_{nlm}$ of a circular Rydberg state of
a principal quantum number $n$ is given by
\begin{eqnarray}
&&
\psi_{n n-1 \pm(n-1)}(\vec{r}) \propto e^{\pm i(n-1)\varphi} 
(r^2 - z^2)^{(n-1)/2}
e^{-r /(na_B)}
\,,
\label{rydberg wave function}
\end{eqnarray}
with the magnetic quantum number $m = \pm (n -1)$, 
and $a_B$ is the Bohr radius.
For large coordinate arguments $r \gg |z|$, 
Rydberg states in general have the radial
wave function of the form,
\begin{eqnarray}
&&
R_{n n-1}(r) \propto r^{n-1}e^{- r/(n a_B)}
\,.
\end{eqnarray}

Momentum representation of Rydberg states is also useful,
as shall be shown in the following.
Appendix B describes momentum space representation of
the wave function.

For subsequent discussion it will become important
to compute a correlation integral of the initial
and the final wave functions of atomic electron,
\begin{eqnarray}
&&
a_{if}(\vec{\Delta}) = 
\int \frac{d^3 x}{(2\pi)^3} e^{-i\vec{\Delta}\cdot\vec{x}}
\psi_f^*(\vec{x})\psi_i (\vec{x})
\,,
\label{correlation integral}
\end{eqnarray}
where $\vec{\Delta} = \vec{q}_1 + \vec{q}_2$ 
is the sum of emitted neutrino momenta.
We assume that both the initial and the final electron
states have definite azimuthal angular momentum components, $m_i \,, m_f$
along the microwave propagation,
taken as $z-$axis.
Using the expansion formula,
\begin{eqnarray}
&&
e^{-i \vec{\Delta}\cdot\vec{x} } = \sum_l i^l (2l + 1) j_l(\Delta r)
P_l (\cos \theta')
\,,
\\ &&
\cos \theta' = \cos \theta_{\Delta} \cos \theta
+ \sin \theta_{\Delta} \sin \theta \cos (\varphi_{\Delta} - \varphi)
\,,
\end{eqnarray}
along with the addition theorem of the spherical harmonics
\begin{eqnarray}
&&
(2l + 1) P_l (\cos \theta') = 4\pi \sum_m Y_{lm}^*(\theta_{\Delta}\,,
\varphi_{\Delta}) Y_{lm}(\theta \,, \varphi)
\,,
\end{eqnarray}
one readily derives $m = m_f - m_i$ after the angular $\varphi$ integration
in eq.(\ref{correlation integral}).

For simplicity we shall work out the correlation integral
only for the transition from a circular to a circular state.
The result is
\begin{eqnarray}
&&
a_{if}(\vec{\Delta}) \approx 
\frac{i^{n_i - n_f}}{\pi\sqrt{2\pi}}c(n_i\,,n_f) 
{\cal P}_{ n_i - n_f}^{n_f - n_i}(\cos \theta_{\Delta})
e^{i (n_f - n_i)\varphi_{\Delta}}
\nonumber \\ &&
\times
\int_{0}^{\infty}dr \,
r^2 j_{n_i - n_f}(\Delta r)R_{n_f n_f-1}(r)R_{n_i n_i-1}(r)
\,,
\\ &&
c(n_i\,,n_f) =
\int_{-1}^1 d\cos \theta \,{\cal P}_{n_f - n_i}^{n_f - n_i}(\cos \theta)
\,{\cal P}_{n_f - 1}^{n_f - 1}(\cos \theta)
\,{\cal P}_{n_i - 1}^{n_i - 1}(\cos \theta)
\,,
\end{eqnarray}
where ${\cal P}_n^m (x)$ is the normalized, associated Legendre polynomial.
We have taken the leading term in $l$ sum, $l = |n_i - n_f|$.
In the large $n$ limit of the principal quantum number,
\begin{eqnarray}
&&
c(n_i\,,n_f) \sim
\pi^{-1/4}
(\frac{n_i}{n_f})^{1/4}
\left( \frac{\Gamma(\,(n_i - n_f + 3)/2}{\Gamma(\,(n_i - n_f + 2)/2}
\right)^{1/2}
\,,
\end{eqnarray}
which is of order unity.
The remaining radial integral is neither small for
the region of
\begin{eqnarray}
&&
\Delta \leq \frac{2\sqrt{2}\pi}{a_B}\frac{1}{n^{3/2}}
\,, \hspace{0.5cm}
\delta n \sim \sqrt{\frac{n}{8}} 
\,,
\end{eqnarray}
with $\delta n = n_i - n_f \ll n_{i\,,f}$.
For the transition around $n \approx 20$,
\begin{eqnarray}
&&
E_i - E_f \sim \frac{2\alpha^2 m_e \delta n}{n^3} \sim 11 meV (\frac{20}{n})^{5/2}
\,,
\\ &&
\Delta \leq \Delta_{max}
\,, \hspace{0.5cm}
\Delta_{max} = 370 eV (\frac{20}{n})^{3/2}
\,.
\label{delta max}
\end{eqnarray}
Thus, the correlation is large for a range of $\Delta \gg E_i - E_f$.

\vspace{0.5cm}
\subsection{Field assisted pair decay}

\hspace*{0.5cm}

Suppose that strong microwave is irradiated to a
Rydberg state $| i \rangle$, which then decays into
a neutrino pair $+ $ another Rydberg state $| f \rangle $;
\begin{eqnarray}
&&
R_i \rightarrow R_f + \nu_i \nu_j
\,.
\end{eqnarray}
Electronic transition $| i \rangle \rightarrow | f \rangle $
under a strong EM field
may be dealt with adopting quantum mechanical treatment of
multiphoton processes \cite{atomic physics}, or its
extention. 
One needs a theoretical formalism
in order to properly incorporate strong field effects.

We consider a time dependent EM field 
in the Hamiltonian taking the radiation gauge,
\begin{eqnarray}
&&
H_1(t) = e\frac{\vec{A}(t)\cdot \vec{p}}{m_e} + e^2\frac{\vec{A}^2(t)}{2m_e}
\,.
\end{eqnarray}
The plane-wave microwave of a linear polarization is given by
the vector potential of the form,
\begin{eqnarray}
&&
\vec{A}(t) = \frac{\vec{E}_0}{\omega}\sin (\omega t)
\,,
\end{eqnarray}
while a circularly polarized case is
\begin{eqnarray}
&&
\vec{A}(t) = \frac{E_0}{\sqrt{2} \omega}(\sin (\omega t) \vec{e}_x \pm 
 \cos (\omega t) \vec{e}_y)
\,,
\end{eqnarray}
where $\vec{e}_i$ is the orthonormal unit vector along $i-$axis.
We took $z-$axis as the direction of light propagation.
In both cases of the polarization $E_0$ is the rms amplitude.

In discussions that follow it is important to
distinguish whether the field is in the strong or the weak
range of the strength.
This may be characterized by interaction strength relative
to its frequency;
\begin{eqnarray}
&&
\frac{eE_0 p}{m_e \omega^2} \sim 
8 \times 10^3 n^{-1} \frac{E_0}{V cm^{-1}}(\frac{GHz}{\omega})^2
\,.
\end{eqnarray}
Depending on a combination of parameters,
$E_0/(\omega^2 n)$, this can be either in the strong or
the weak field range.
Even if the field strength is strong in this sense,
it may be arranged that the field is not large enough to
ionize Rydberg electron, $eE_0 < \alpha /(n^2 a_B)$
(attractive force from nucleus),
which means
\begin{eqnarray}
&&
E_0 < 7 \times 10^{11}  n^{-2}V cm^{-1}
\,.
\end{eqnarray}

We would like to treat the weak interaction process alone perturbatively,
and solve interaction of atomic electrons with microwave
as analytically as possible.
As discussed in Appendix B,
the transition amplitude of the neutrino pair
emission is given by
\begin{eqnarray}
&&
(S - 1)_{fi} \sim - i \int \frac{d^3 q_1 d^3 q_2}{(2\pi)^6}
\int \frac{d^3 p_f d^3 p_i}{(2\pi)^3} \times
\nonumber \\ &&
\hspace*{-1cm}
\int_{- \infty}^{\infty}\, dt_1\,
\langle A_f(\infty)|\vec{p}_f \rangle 
\langle \vec{p}_f |
U_A(\infty \,,t_1)H_W U_A(t_1\,, - \infty)|\vec{p}_i \rangle 
\langle \vec{p}_i | A_i(- \infty)\rangle
\label{s-matrix 2}
\,.
\end{eqnarray}
Here $H_W$ is the weak interaction Hamiltonian of the neutrino 
pair emission of momenta $\vec{q}_i$.

States taken as initial and final ones in eq.(\ref{s-matrix 2}), 
$U_A(t\,, - \infty)| A_i(- \infty)\rangle$ and 
$U_A(t\,, \infty )| A_i( \infty)\rangle$ 
are
bound state solutions of the Schroedinger equatioin governed by
the Hamiltonian of Coulomb potential plus the microwave field.
Their momentum space representation $U_A(t\,, - \infty)|\vec{p}_i \rangle $ 
and $U_A(t\,, \infty ) |\vec{p}_f \rangle$ are used here.
The approximation, taken by Keldysh \cite{keldysh}
 and the one we shall also adopt,
is to neglect the Coulomb interaction during
the occurrence of weak process, and use the plane-wave solution
under the periodic field, known as
the Volkov solution \cite{atomic physics}; 
the time dependent part in this approximation is
\begin{eqnarray}
&&
\langle \vec{p}_f | U_A(\infty \,, t) |\vec{p}_f \rangle 
\langle \vec{p}_i |U_A(t\,, - \infty)| \vec{p}_i \rangle
\propto
\nonumber \\ &&
\exp [-i \left(\frac{\vec{p}_i^2 - \vec{p}_f^2}{2m_e}\,t + 
\frac{(\vec{p}_i - \vec{p}_f)\cdot \int_{-\infty}^{t} dt_1 \vec{A}(t_1)}{m_e}
\,\right)]
\,,
\nonumber
\end{eqnarray}
negelcting irrelevant phase factors.

The Keldysh approximation is valid if correction of the binding
is small.  This condition is worked out in \cite{reiss},
and in our case it leads to the frequency condition,
as discussed in Appendix B,
\begin{eqnarray}
&&
\omega \leq 2.1 \times 10^6 n^{-4} GHz
\,,
\end{eqnarray}
with $n$ the maximal principal quantum number during
the transition.
The field strength is also limited as in Appendix B.

The weak Hamiltonian $H_W$ of neutrino pair emission
is translationally invariant,
hence its matrix element contains the momentum conserving
delta function;
\begin{eqnarray}
&&
\hspace*{1cm}
\langle \vec{p}_f |
U_A(\infty \,,t)H_W U_A(t\,, - \infty)|\vec{p}_i \rangle 
=
\nonumber \\ &&
\exp [-i \left((\frac{p_i^2}{2m_e} - \frac{p_f^2}{2m_e}
- E_1 - E_2)\,t + 
\frac{(\vec{p}_i - \vec{p}_f)\cdot\int_{-\infty}^{t} dt_1\vec{A}(t_1)}{m_e }
\,\right)]
\nonumber \\ &&
\times
\frac{G_F}{\sqrt{2}}(2\pi)^3 
\delta (\vec{p}_i - \vec{p}_f - \vec{q}_1 - \vec{q}_2) 
\sum_{ij} \langle f_{\nu}|j_{ij}|0 \rangle \cdot
\langle f | j_{ij}^e | i \rangle
\,.
\end{eqnarray}
The neutrino pair is not observed, hence one takes 
the helicity and momentum summation of neutrinos.
The helicity summation of
\begin{eqnarray*}
&&
\sum_{hh'}\sum_{ij} |\langle f_{\nu}|j_{ij}|0 \rangle \cdot
\langle f | j_{ij}^e | i \rangle|^2
\,,
\end{eqnarray*}
has been examined in Section \lromn3 in
detail, yielding in the leading approximation of $1/m_e$, 
the marix element squared of the form, $(2 - m_1 m_2 /(E_1 E_2))$,
eq.(\ref{majorana h-sum}), times the electron wave function factors.

To proceed further for computation of the transition amplitude
squared, we insert a convenient identity, 
\begin{eqnarray}
&&
\int d\vec{\Delta} \int dE_{12} \delta( \vec{\Delta} - \vec{q}_1 - \vec{q}_2 ) 
\delta (E_{12} - E_1 - E_2)
= 1
\,,
\nonumber
\end{eqnarray}
with $E_i = \sqrt{q_i^2 + m_i^2}$ neutrino momenta,
and use
\begin{eqnarray}
&&
|\int \frac{d^3 q_1 d^3 q_2}{(2\pi)^6}
\int d\vec{\Delta} \int dE_{12} (2\pi)^3\delta( \vec{\Delta} - \vec{q}_1 - \vec{q}_2 ) 
\delta (E_{12} - E_1 - E_2) f(\vec{q}_1 \,, \vec{q}_2)|^2
\nonumber \\ &&
\hspace*{-1cm}
= \int d\vec{\Delta} \int dE_{12} 
\int \frac{d^3 q_1 d^3 q_2}{(2\pi)^3}
\delta( \vec{\Delta} - \vec{q}_1 - \vec{q}_2 ) 
\delta (E_{12} - E_1 - E_2) |f(\vec{q}_1 \,, \vec{q}_2)|^2
\,.
\end{eqnarray}
Neutrino momentum integration here is
\begin{eqnarray}
&&
\hspace*{-1.5cm}
K_{ij}^{M}(E_{12} \,, \vec{\Delta}) =
\int \frac{d^3 q_1 d^3 q_2}{(2\pi)^3}
\delta(\vec{\Delta} - \vec{q}_1 - \vec{q}_2) \delta(E_{12} - E_1 - E_2) 
(2 - \frac{m_1 m_2}{E_1 E_2})
\\ &&
= \frac{1}{(2\pi)^2}\sqrt{\left(
1 - \frac{(m_1 + m_2)^2}{s_{12}}\right)
\left(1 - \frac{(m_1 - m_2)^2}{s_{12}})
\right)}
\nonumber \\ &&
\times
\left(
E_{12}^2 (1 + \frac{m_1^2 - m_2^2}{s_{12}})
- \frac{E_{12}^2}{2}(1 + \frac{m_1^2 - m_2^2}{s_{12}})^2
- m_1 m_2
\right.
\nonumber \\ &&
\left.
- \frac{\vec{\Delta}^2}{6}
(1 - \frac{(m_1 + m_2)^2}{s_{12}})
(1 - \frac{(m_1 - m_2)^2}{s_{12}})
\right)
\,,
\hspace{0.5cm}
s_{12} = E_{12}^2 - \vec{\Delta}^2
\,.
\end{eqnarray}
In the massless neutrino limit, 
\begin{eqnarray}
&&
K_{ij}^{M}(E_{12} \,, \vec{\Delta})
\rightarrow
\frac{1}{8\pi^2} (E_{12}^2 - \frac{\vec{\Delta}^2}{3} )
\,.
\label{massless case of ph}
\end{eqnarray}
The threshold behavior of this quantity at $s_{12} \rightarrow (m_1 + m_2)^2$ 
is
\begin{eqnarray}
&&
K_{ij}^{M}(E_{12} \,, \vec{\Delta})
\sim
\frac{(m_1 m_2)^{3/2}}{2\pi^2 (m_1 + m_2)^2}
\sqrt{E_{12}^2 - \vec{\Delta}^2 - (m_1 + m_2)^2}
\,.
\label{threshold k}
\end{eqnarray}

The transition probability is further simplified first by
using 
\begin{eqnarray}
&&
\hspace*{-1cm}
\langle \vec{p}_i | e^{-i p_i^2 t/ (2m_e)} | A_i(- \infty) \rangle
= \langle \vec{p}_i | e^{-i H_0 t} | A_i(- \infty) \rangle
= e^{- iE_i t} \langle \vec{p}_i |  A_i(- \infty) \rangle
\,,
\end{eqnarray}
and a similar relation for the final state, to
replace the time dependent factor to $e^{-i (E_i - E_f) t}$.
When the Fourier transformation back to
the configuration space is made, one obtains
\begin{eqnarray}
&&
\int \frac{d^3 p_f d^3 p_i}{(2\pi)^3}
\delta (\vec{p}_i - \vec{p}_f - \vec{\Delta})
\langle A_f(\infty)|\vec{p}_f \rangle 
\langle \vec{p}_i | A_i(- \infty)\rangle
\nonumber \\ &&
= 
\int \frac{d^3 x}{(2\pi)^3} 
e^{i\vec{\Delta}\cdot\vec{x}} \psi_{f}^*(\vec{x})\psi_{i}(\vec{x})
\equiv a_{if}(\vec{\Delta})
\,.
\end{eqnarray}
The transition probability is then
\begin{eqnarray}
&&
|(S - 1)_{fi}|^2 = 
\frac{ G_F^2}{2}
\int d\vec{\Delta} |a_{if}(\vec{\Delta})|^2
\int_{\sqrt{\vec{\Delta}^2 + (m_1 + m_2)^2}}^{\infty} dE_{12} 
\, K_{ij}^{M}(E_{12} \,, \vec{\Delta}) 
\nonumber \\ &&
\hspace*{1cm}
\times
|\int_{- \infty}^{\infty}\, dt\,{\cal F}_{if}(E_{12} \,, \vec{\Delta} \,;t)|^2
\,,
\label{master formula}
\\ &&
{\cal F}_{if}(E_{12} \,, \vec{\Delta} \,;t) 
= \exp [-i \left(\Delta_{if}t  + \int_{-\infty}^{t}dt_1
\frac{e\vec{\Delta}\cdot\vec{A}(t_1)}{m_e }
\,\right)]
\,,
\\ &&
\hspace*{1cm}
\Delta_{if} = E_i - E_f - E_{12}
\,.
\label{energy difference}
\end{eqnarray}

In the formula (\ref{master formula}) three important factors are separated;
the inegrated neutrino factor $K_{ij}^M$, 
the initial and the final electron factor $|a_{ij}(\vec{\Delta})|^2$, and
the time dependence factor ${\cal F}_{if}$ related to microwave irradiation.
This essential simplification owes to
the Keldysh approximation.

The time integral for this S-matrix element invloves
the time interval of infinite duration.
In practice, it is important to understand a finite
time integral of the form,
\begin{eqnarray}
&&
\hspace*{-1cm}
\int_{t_0}^{t}\, dt_1\,{\cal F}_{if}(E_{12} \,, \vec{\Delta}
\,;t_1)
= \int_{t_0}^{t}\, dt_1\,
\exp [-i \left(\Delta_{if}t_1 + \int_{-\infty}^{t_1}dt_2
\frac{e\vec{\Delta}\cdot\vec{A}(t_2)}{m_e}
\,\right)]
\,.
\label{time dependence}
\end{eqnarray}

The analysis of this time dependent phenomena
is separated into two parts;
in the first part we present the conventional multiphoton
picture, which corresponds to the narrow band region
of a more general analysis in the last subsection.
There exists the additional wide band region which typically
exhibits the exponentially growing instability,
a phenomenon very time dependent.
Only a lower limit of the growing rate is estimated in the
wide band region.
The real process goes with these two mechanisms entangled,
hence is complicated.

\vspace{0.5cm} 
\subsection{Multiphoton picture}

\hspace*{0.5cm}

We postpone a general analysis of eq.(\ref{time dependence}), and
concentrate here on an approximate expansion that leads to
interpretation based on multiphoton processes.
The multiphoton process has been described in many
textbooks; for instance in \cite{atomic physics} for strong laser field,
in \cite{rydberg atom} and \cite{cohen-tannoudji} for microwave or rf fields.

We shall first discuss the case of linear microwave polarization.
Using the expansion in terms of the Bessel function $J_N(x)$,
\begin{eqnarray}
&&
e^{- ib \sin \omega t} = \sum_{N= -\infty}^{\infty} 
J_N(b)e^{- iN\omega t}
\,, \hspace{0.5cm}
b = \frac{e\vec{\Delta}\cdot \vec{E}_0 }{m_e \omega^2}
\,.
\end{eqnarray}
eq.(\ref{time dependence}) becomes a sum of simple exponentials.
The time integral is then readily computed,
leading in the large time limit to 
\begin{eqnarray}
&&
\hspace*{-1cm}
\int_{t_0}^{t}\, dt_1\,
{\cal F}_{if}(E_{12} \,, \vec{p}_i \,, \vec{p}_f \,;t_1) 
\rightarrow 
2\pi \sum_N
J_N(b) \delta(\Delta_{if} + N\omega)
\,.
\end{eqnarray}
The argument of the delta function implies
the simple energy conservation, due to (\ref{energy difference}).

It would be instructive to numerically estimate 
the important quantity that appears in these formulas;
the magnitude of microwave interaction,
\begin{eqnarray}
&&
b \sim \frac{eE_0 \Delta }{m_e\omega^2}
\approx 0.2 \, \frac{\Delta}{0.1 eV}\frac{E_0}{V cm^{-1}}(\frac{GHz}{\omega})^2
\,.
\end{eqnarray}
The momentum scale $\Delta$ has been set here
to around 2 times the neutrino mass.
This parameter $b$ can be very large.
The Bessel function $J_N(b)$ is maximal at $b \approx |N|$.

Using the standard formula,
\begin{eqnarray*}
\lim_{t \rightarrow \infty} \frac{|2\pi \delta (\Delta)|^2}{t} = 
2\pi \delta (\Delta)
\,,
\end{eqnarray*}
we derive 
a time independent rate $w = \lim_{t \rightarrow \infty}P(t)/t$ 
of the form,
\begin{eqnarray}
&&
w = \sum_N w_N
\,, \hspace{0.5cm}
w_N = 
\pi G_F^2
\int d\vec{\Delta} 
|a_{if}(\vec{\Delta})|^2
|J_{N}\left(\frac{e\vec{\Delta} \cdot\vec{E}_0}{m_e \omega^2}\right)|^2
\nonumber \\ &&
\times
\int_{\sqrt{\vec{\Delta}^2 + (m_1 + m_2)^2}}^{\infty} dE_{12}
\delta ( E_{12} - N\omega - E_i + E_f )
K_{ij}^{M}(E_{12} \,, \vec{\Delta})
\,.
\label{rate formula 0}
\end{eqnarray}
There is a minimum number of photons $N_0$ in the $N$ summation;
$N > N_0$ with 
\begin{eqnarray}
&&
N_0 = \frac{1}{\omega} \left(
\sqrt{\vec{\Delta}^2 + (m_1 + m_2)^2}  - E_i + E_f 
\right)
\,.
\end{eqnarray}
This $N_0$ can be negative.
A negative $N_0$ means that stimulated microwave photons are  
emitted (instead of absorbed) to cause the neutrino pair emission.
The requirement for a negative $N_0$ is
\begin{eqnarray}
&&
E_i - E_f > m_1 + m_2
\,.
\end{eqnarray}
For the weak field the neutrino pair emission rate is in proportion to
\begin{eqnarray}
&&
|J_{N}\left(\frac{e\vec{\Delta}\cdot \vec{E}_0 }{m_e \omega^2}
\right)|^2 
\propto (\frac{E_0}{\omega})^{2|N|}
\,.
\end{eqnarray}

Interpretation of this result is that there are
contributions from $N$ photon absorption for $N >0$ or $-N$
emission for $N<0$, which feed or take away energy $|N|\omega$
to cause the multiphoton transition.
Roughly, the relation to the neutrino mass $m_i$,
$m_1 + m_2 \approx E_i - E_f + N\omega$  holds.
Suppose that the final state is specified.
When $E_f > E_i$ (the case of upper level), 
only microwave absorption is possible for the pair emission.
When $E_f < E_i$ (the case of lower level), 
both absorption and emission is possible according to
the sign of $m_1 + m_2 - (E_i - E_f)$.
Since the rate is maximal at $|N| \approx b \propto E_0$,
the adjustment of the field amplitude $E_0$ can help
to locate the position of the threshold of
$E_i - E_f = m_1 + m_2$.

The neutrino pair emission accompanying $N (>0)$ microwave absorption
occurs as if a hypothetical heavy 'boson'
of mass $ N\omega + E_i - E_f$ decays according to the rate
$w_N$.
Let us clarify this process in the weak field limit.
The neutrino pair emission caused by $N$ multiphoton transition 
occurs with interaction strength,
\begin{eqnarray}
&&
\frac{G_F}{\sqrt{2}}\frac{1}{N!}
(\frac{e\vec{\Delta}\cdot\vec{E}_0}{2m_e \omega})^{N}
\,.
\end{eqnarray}
This is the E1 transition repeated $N$ times plus the weak
pair emission.
The hypothetical 'boson' does not have a definite
momentum to be transimitted to the neutrino pair, but
the conservation law is replaced by
\begin{eqnarray*}
&&
(2\pi)^3 \delta (\vec{\Delta})
\rightarrow |a_{if}(\vec{\Delta})|^2
\,.
\end{eqnarray*}
There is nothing special about this, because
both initial and final states are not momentum eigenstates.
The momentum distribution has a width $\approx \alpha^2 m_e/n^3$.
The total rate is a sum over $N$ of many multiphotons.
A large mass 'particle' of $N \gg 1$ might be called a heavy electron due
to many photon clouds.
The rate of a very heavy electron decay is suppressed by
\begin{eqnarray}
&&
(\frac{1}{N!})^2
(\frac{E_i - E_f - E_{12}}{2\omega })^{2N}
\,,
\end{eqnarray}
because the Bessel function behaves as $|J_N(b)| \sim (b/2)^{2N}/(N!)^2$.

To obtain a large rate, it is necessary to have
many contributions of different $N$.
The property of the Bessel function $J_N(b)$ tells that
the $N$ sum can be large when $N$ is of order $b$ or less.
The ratio $b/N$, in particular
\begin{eqnarray}
&&
\frac{b}{|N_0|} = \frac{e\vec{\Delta}\cdot\vec{E}_0}
{\sqrt{\Delta^2 + (m_1 + m_2)^2} m_e \omega }
\,,
\end{eqnarray}
can however be small, especially
if $\Delta \gg m_1 + m_2$, because in this case
\begin{eqnarray}
&&
\frac{b}{|N_0|} = O[\frac{eE_0}{m_e \omega}]
\sim O[0.9 \times 10^{-5}]
\frac{E_0}{V cm^{-1}}\frac{GHz}{\omega}
\,.
\end{eqnarray}
Thus, it is important to have for $N_0$ a small,  or better,
a negative number such that the $N$ sum contains
contributions from small $N$'s.

A better, but still crude way of estimating a number of multiphoton
contributions of many paths, hence the pair emission rate is as follows.
Take $\vec{E}_0$ along $z-$direction.
For a given $\Delta_z$,
there is a region of relatively large value of the Bessel 
function $J_N(b)$ at $b \approx N \gg 1$,
\begin{eqnarray}
&&
J_N(\frac{e \Delta_z E_0}{m_e \omega^2})
\sim c (\frac{m_e \omega^2}{e \Delta_z E_0})^{1/3}
\,, \hspace{0.5cm}
c = \frac{\Gamma(1/3)}{2^{2/3}3^{1/6}\pi} \sim 0.45
\,.
\end{eqnarray}
Replacing the Bessel fuction by this gives the rate,
\begin{eqnarray}
&&
\hspace*{-1cm}
w = \pi c^2 G_F^2 \int d\vec{\Delta}
K_{ij}^{M}(\frac{e \Delta_z E_0}{m_e \omega} + E_i  - E_f 
\,, \vec{\Delta})|a_{if}(\vec{\Delta}))|^2
(\frac{m_e \omega^2}{e \Delta_z E_0})^{2/3}
\,.
\label{rate formula 1}
\end{eqnarray}

The threshold behavior of the rate based on
(\ref{rate formula 1}) may be derived using 
the neutrino factor $K_{ij}^M$ near the threshold, (\ref{threshold k}).
For this estimate we assume the correlation integral of
order unity $a_{ij} = O[1]$, to give
\begin{eqnarray}
&&
\hspace*{-1.5cm}
c^2G_F^2 J \frac{(m_1 m_2)^{3/2}}{(m_1 + m_2)^2}
(\frac{m_e \omega^2}{eE_0})^{2/3}
\sum_N
\left( (N\omega + E_i - E_f)^2 - (m_1 + m_2)^2\right)^{5/3}
\,,
\nonumber \\ &&
\end{eqnarray}
where $J$ is 
\begin{eqnarray}
&&
J = \int_0^1 d\rho \rho (1 - \rho^2)^{5/6}\int_0^1 dz z^{-2/3} 
\sqrt{1 - z^2}
\,, \hspace{0.5cm}
c^2 J \sim 0.15
\,.
\end{eqnarray}
The $N$ summation is limited by 
\begin{eqnarray}
&&
\hspace*{-1cm}
N < N_{max}
\,, \hspace{0.5cm}
N_{max} = 
O[\frac{eE_0 \Delta_{if}}{m_e \omega^2}]
\sim 0.2 \frac{E_0}{V cm^{-1}}\frac{m_{\nu}}{50 meV}(\frac{GHz}{\omega})^2
\,,
\end{eqnarray}
where $\Delta_{if}$ is a typical momentum transfer of order
$m_1 + m_2$.

The enhancement factor $R$ may be defined relative to
the standard rate near the threshold,
$G_F^2 (m_1 + m_2)^5/(15\pi^3)$;
\begin{eqnarray}
&&
R \sim \frac{45 \pi^3 c^2 J}{16}(\frac{m_e \omega^2}{eE_0 })^{5/3}
\frac{(m_1 m_2)^{3/2}}{\omega^2 \Delta_{if} (m_1 + m_2)^7}
\nonumber \\ &&
\times
\left( (\frac{eE_0\Delta_{if}}{m_e \omega} + E_i - E_f)^2 
- (m_1 + m_2)^2\right)^{8/3}
\,.
\end{eqnarray}
For the weak field of $eE_0/(m_e \omega) \ll 1$, the threshold appears at
$E_f = E_i - m_1 - m_2$.
It may also appear as a threshold of the field amplitude $E_0$.
A large power $8/3$ implies that the rate quickly increases
once the threshold is passed.

Well above the threshold one may use eq.(\ref{massless case of ph}),
and the rate and the enhancement factor become of order,
\begin{eqnarray}
&&
\frac{27 c^2 G_F^5 \omega^5}{448} (\frac{m_e \omega}{eE_0})^{2/3}
( N_{max} + \frac{E_i - E_f}{\omega})^{16/3}
\,, 
\\ &&
R \sim  5.6 (\frac{eE_0}{m_e \omega})^{14/3}
(\frac{m_1 + m_2}{\omega})^{1/3}
\,,
\end{eqnarray}
taking $\Delta_{if} = m_1 + m_2$.
This factor is very sensitive to the microwave power $\propto P^{7/3}$
and its frequency $\propto \omega^{-5}$, and
its precise determination requires a more elaborate computation.

There is also contribution from $N < b$ region of the Bessel
function $J_N(b) \sim \sqrt{\frac{\pi}{2b}}$.
This contribution is estimated as
\begin{eqnarray}
&&
w_{N< b} = O[\frac{\pi G_F^2 (\Delta E)^5}{240}\frac{m_e \omega}{eE_0}]
\,.
\end{eqnarray}

We shall finally consider the case of circular polarization.
In this case
\begin{eqnarray}
&&
\vec{\Delta}\cdot\vec{A}(t) = \frac{E_0 \Delta}{\sqrt{2}\omega^2}
\sin \theta_{\Delta} \sin (\omega t \pm \varphi_{\Delta})
\,,
\end{eqnarray}
$\theta_{\Delta}\,, \varphi_{\Delta}$ being
angle factors of the momentum $\vec{\Delta}$.
The expansion in terms of the Bessel function becomes
\begin{eqnarray}
&&
\hspace*{-1cm}
\exp [i e\frac{\vec{\Delta}\cdot\vec{A}(t)}{m_e}] =
\sum_{N_d} 
J_{N_d}\left( \frac{eE_0 \Delta \sin \theta_{\Delta}}{\sqrt{2}m_e\omega^2} 
\right)
\exp[-i N_d \omega t \mp i  N_d \varphi_{\Delta})]
\,.
\end{eqnarray}
The pair emission rate for a circular to a circular transition is given
in terms of the correlation integral by
\begin{eqnarray}
&&
w = 2\pi^2 G_F^2 \int_0^{\infty}d\Delta \Delta^2 \int_{-1}^1 
d\cos \theta_{\Delta}
|a_{if}(\Delta)|^2 {\cal P}_{n_f + n_i - 1}^{n_f - n_i}(\cos \theta_{\Delta}) 
\times
\nonumber \\ &&
\theta \left( E_i - E_f \mp (n_i - n_f) \omega  -
\sqrt{\Delta^2 + (m_1 + m_2)^2} \right) \times
\nonumber \\ &&
|J_{ n_i - n_f}\left( \frac{eE_0 \Delta \sin \theta_{\Delta}}
{\sqrt{2}m_e\omega^2} \right)|^2
K_{ij}^{M}(E_i \mp (n_i - n_f) \omega - E_f \,, \Delta)
\,.
\end{eqnarray}

A detailed and more precise rate computation in the
multiphoton picture shall be presented elsewhere,
and be compared to a more general approach presented
in the following subsection.
The method described in the present subsection has a limitation, 
and is interpreted
as a part of more general approach we shall now describe.

\vspace{0.5cm} 
\subsection{Relevance of parametric resonance}

\hspace*{0.5cm}
Consider the time dependent part of the rate, 
\begin{eqnarray}
&&
{\cal W}_{if}(t\,; E_{12}\,, \vec{\Delta})
= 
{\mathrm{Re}} \left(
{\cal F}^*_{if}(t)
\int_{t_0}^{t}\, dt_1\,{\cal F}_{if}(t_1)
\right)
\,,
\label{rate 0}
\end{eqnarray}
which appears in the transition rate given by
the time derivative of the transition probability (\ref{master formula}),
namely
\begin{eqnarray}
&&
\hspace*{-1cm}
w(t) =
\frac{ G_F^2}{2}
\int d\vec{\Delta} |a_{if}(\vec{\Delta})|^2
\int_{\sqrt{\vec{\Delta}^2 + (m_1 + m_2)^2}}^{\infty} dE_{12} 
\, K_{ij}^{M}(E_{12} \,, \vec{\Delta}) 
{\cal W}_{if}(t\,;E_{12}\,, \vec{\Delta})
\,.
\nonumber \\ &&
\end{eqnarray}
We define a complex function $G(t\,; a\,, b)$ 
in terms of new dimensionless variables $a\,, b$
\begin{eqnarray}
&&
G(t\,; a\,, b) = e^{ia \omega t + ib \sin \omega t}
\int_0^t dt'\, 
e^{- ia \omega t' - ib \sin \omega t'} \,,
\\ &&
a = \frac{E_i - E_f - E_{12}}{\omega}
\,, \hspace{0.5cm}
b =  \frac{e \vec{\Delta}\cdot\vec{E}_0}{m_e \omega^2}
\,,
\label{time dependent rate 0}
\end{eqnarray}
such that 
${\cal W}_{if}(t\,;E_{12}\,, \vec{\Delta}) 
= G(t\,; a(E_{12})\,, b(\vec{\Delta})\,)$.

The multiphoton picture in the preceeding subsection
corresponds to an infinite time limit ignoring a coherence
in the computation of $G(t\,; a\,, b)$, which leads to a
time independent, constant rate $w(\infty)$.
However, there exists an intrinsic instability in
some parameter region of $(a\,, b)$, which we now discuss.
A coherence effect at finite times is crucial in this discussion.

The quantity (\ref{time dependent rate 0}) 
satisfies a coupled differential equation,
\begin{eqnarray}
&&
\hspace*{-1.5cm}
\left(
\begin{array}{cc}
\frac{d^2}{dt^2} + \omega^2 (a + b  \cos \omega t)^2 
&  -b \omega^2  \sin \omega t \\
b \omega^2  \sin \omega t 
& \frac{d^2}{dt^2} + \omega^2 (a + b \cos \omega t)^2
\end{array}
\right)
\left(
\begin{array}{c}
  {\mathrm{Re}} G(t) \\
  {\mathrm{Im}} G(t)
\end{array}
\right)
= 0
\,.
\label{differential eq for g}
\end{eqnarray}
This system has an intrinsic frequency scale $|a|\omega
= |E_i - E_f - E_{12}|$
such as given by the energy difference between the initial
and the final states.
This intrisic scale is further modulated by a periodic variation of 
parameters whose frequency is $\omega$
and amplitude is $|b|$.
Cooperative effects of external modulation with the intrinsic
property gives rise to interesting phenomena of the 
parametric amplification.

In our time dependent problem, the initial condition is specified as
\begin{eqnarray}
&&
G(0) = 0 \,, \hspace{0.5cm}
\dot{G}(0) = 1
\end{eqnarray}
Thus, there is no way to avoid the instability of the
parametric resonance, once the parameters $(b\,,a)$ fall in
the instability band.

General theory \cite{landau-lifschitz m} 
of linear differential equations with periodic
coefficients indicates solutions of the Mathieu type,
and the unstable and stable band structure appears
in the parameter $(b\,,a)$ plane.
In the unstable band the  exponential growth is observed;
\begin{eqnarray}
&&
{\mathrm{Re}} 
\left( e^{\Gamma(a\,, b)t/2 - i (E_i - E_f - E_{12})t} f(\omega t) \right)
\,,
\end{eqnarray}
with $f(\tau)$ a periodic function,
hence the instability greatly expediates depletion of the
prepared state.
The parameter $b$ is essentially $\propto$ the total momentum of
the neutrino pair $ \Delta_z$ projected onto
the microwave electric field direction taken $z-$ axis here, which is
also a typical momentum transfer in the electron transition
$|i \rangle \rightarrow |f \rangle$.
Another one $a \propto E_{12} - E_i + E_f $ is the
total neutrino energy minus the mass difference
$E_i - E_f$.
Thus, $(b\,,a) \propto (\Delta_z \,, E_{12} - E_i + E_f)$,
and the unstable band structure in $(b\,,a)$ plane
signifies where in the phase space of the neutrino pair 
contributes to the emission rate.
The instability signifies an exponential decay
of the initially prepared state.

The relative weight of $b$ term in eq.(\ref{differential eq for g}),
the magnitude $|b|/|a|$, signifies the importance of
the parametric amplification. 
Roughly, the narrow band region is in the parameter
region of $|b| \ll |a| $, and the wide band region
is in $|b| \gg |a| $.
The multiphoton picture explained in the previous
subsection corresponds to a narrow band of instability.
The $N-$th band in the narrow band region corresponds to a  mass difference
of initial and final states,
$N\omega + E_i - E_f $ shifted by the energy input of $N$
microwave photons.
The narrowness implies weaker rates.
The diagramatic interpretation of the narrow band decay has been given
in the literature \cite{narrow band interpretation}.
The correspondance between the two regions is given by
\begin{eqnarray}
&&
\hspace*{-1.5cm}
\sum_N |J_N(\frac{e\vec{\Delta}\cdot\vec{E}_0}{m_e \omega^2})|^2 2\pi 
\delta (N\omega + E_i - E_f - E_{12} )
\; \leftrightarrow \; 
{\cal W}_{if}(t\,;E_{12}\,, \vec{\Delta})
\,.
\label{wide resonance}
\end{eqnarray}
Rather than a discrete $N$ sum of multiphotons there is
a continuous spectrum of heavy electrons of mass $E_{12}$
present in the wide band region.

The wide band region gives a much more enhanced 
time dependent rate than the
narrow band multiphoton result presented in the previous
subsection.
If $(b\,,a)$ lies deeper in the instability band, namely,
the larger $|b|/ |a|$ is, the greater the rate is. 
The rate readily exceeds order unity (in the unit
of $1/\omega$, that is $\Gamma = O[\omega]$) deep in an
instability band.
In the wide band region there is no definite number of muliphotons $N$,
and the continuum broad mass $'E_i' - E_f$ range all contribute to
the instability.
If one experimentally arranges that there is no radiatively decaying state
as in the cavity QED, then this means that the neutrino pair
emission is expediated; enhanced pair emission.
It is thus important to depict the structure of stability-instability
bands in $(b\,,a)$ plane.

The phase space region in terms of $(E_{12}\,, \vec{\Delta})$
of the wide band region is estimated as follows.
One can imagine that the most important region is restricted to
$|a| \leq |b|$ due to an experience in the Mathieu equation.
For large $b$'s, the coupled equation (\ref{differential eq for g}) 
approximately decouples,
and the coefficient term of the form, $\propto
(1 - \cos 2\omega t)$ implies that the relevant $(b\,, a)$ region is
indeed deep in wide band regions.
The exponential growth rate given by
$e^{\lambda \omega t}/(\lambda\omega)$
is of order $\lambda \approx 0.15/2$, as shown in \cite{mine95-96-2}. 

The two constraints on the wide band region $|a| \leq |b|$ and the 
mass restriction correspond to a region of $(E_{12}\,, \vec{\Delta})$,
\begin{eqnarray}
&&
(E_{12} -  E_i + E_f)^2 \leq (\frac{eE_0}{m_e \omega})^2 \Delta_z^2
\,, \hspace{0.5cm}
E_{12}^2 \geq \vec{\Delta}^2 + (m_1 + m_2)^2
\,.
\end{eqnarray}
This gives a neutrino pair momentum integration of order,
\begin{eqnarray}
&&
2\int d\vec{\Delta} |a_{if}(\vec{\Delta})|^2
\int_{E_i - E_f - \epsilon |\Delta_z|}^{E_i - E_f + \epsilon |\Delta_z|} dE_{12} 
K_{ij}^{M}(E_{12} \,, \vec{\Delta}) {\cal W}_{if}(t\,;E_{12}\,, \vec{\Delta})
\,,
\end{eqnarray}
for $\epsilon \ll 1$ with
\begin{eqnarray}
\epsilon = |\frac{eE_0}{m_e \omega}| 
\,.
\end{eqnarray}
For a stronger field, the phase space area is quite different.
In particular, towards and above the critical strength $E_c$,
\begin{eqnarray}
&&
E_c = \frac{m_e \omega}{e} \sim 1.1 \times 10^5 \,V cm^{-1} \frac{\omega}{GHz}
\,,
\end{eqnarray}
the momentum space integration is changed to
\begin{eqnarray}
&&
\hspace*{-1cm}
2\int d\vec{\Delta} |a_{if}(\vec{\Delta})|^2
\left(
\theta (\Delta_* - \Delta_z)
\int_{E_i - E_f - \epsilon |\Delta_z|}^{E_i - E_f + \epsilon |\Delta_z|} 
dE_{12} 
\right.
\nonumber \\ &&
\hspace*{-1cm}
\left.
+ \theta (\Delta_z - \Delta_*)
\int_{\sqrt{\Delta^2 + (m_1 + m_2)^2}}^{E_i - E_f + \epsilon |\Delta_z|} 
dE_{12} 
\right)
K_{ij}^{M}(E_{12} \,, \vec{\Delta}) {\cal W}_{if}(t\,;E_{12}\,, \vec{\Delta})
\,.
\end{eqnarray}
In the $\epsilon \gg 1$ limit the second term is dominant, and
\begin{eqnarray*}
\Delta_* \approx \frac{1}{\epsilon}
\left(
E_i - E_f + \sqrt{\Delta_x^2 + \Delta_y^2 + (m_1 + m_2)^2}
\right)
\,.
\end{eqnarray*}

We shall make a crude estimate for the pair emission rate
in the wide band region by making two assumptions.
We first introduce an average rate factor 
$\langle {\cal W}_{if} \rangle$,
and next compute the rate far away from the threshold region for
which one may use eq.(\ref{massless case of ph}).
The $E_{12}$ integration then gives
\begin{eqnarray}
&&
\approx
\int d\vec{\Delta} |a_{if}(\vec{\Delta})|^2
\frac{\epsilon^3}{24\pi^2}|\Delta_z|^3
\langle {\cal W}_{if} \rangle
\,.
\end{eqnarray}
It is thus expected to obtain the rate larger than
\begin{eqnarray}
&&
\hspace*{1cm}
w \approx 
\frac{G_F^2 \Delta_{max}^6}{240\pi}(\frac{eE_0}{m_e \omega})^3
\langle {\cal W}_{if} \rangle
\\ &&
\sim 6 \times 10^{-28} s^{-1} 
\left( \langle {\cal W}_{if} \rangle \omega/10 \right)
(\frac{E_0}{kV cm^{-1}})^{3}
(\frac{GHz}{\omega})^{4}(\frac{\Delta_{max}}{400 eV})^{6}
\,,
\label{rate estimate 0}
\end{eqnarray}
using $\Delta_{max}$ of order, eq.(\ref{delta max}).

Once the large rate is confirmed, one may go to
the threshold region, in which
one replaces the energy integral by
\begin{eqnarray}
&&
\hspace*{1cm}
\int_{\sqrt{\Delta^2 + (m_1 + m_2)^2}}^{E_i - E_f + \epsilon |\Delta_z|} 
dE_{12} K_{ij}^{M}(E_{12} \,, \vec{\Delta}) 
\nonumber \\ &&
\hspace*{-1cm}
\sim
\frac{(m_1 m_2)^{3/2}}{4\pi^2 (m_1 + m_2)^2}
(E_i - E_f + \epsilon |\Delta_z|)
\sqrt{(E_i - E_f + \epsilon |\Delta_z|)^2 - \vec{\Delta}^2 - (m_1 + m_2)^2}
\,.
\nonumber \\ &&
\end{eqnarray}
This gives the threshold rate, with $\epsilon = eE_0/(m_e \omega)$,
\begin{eqnarray}
&&
\hspace*{-1.5cm}
\frac{G_F^2 \Delta_{max}^2}{6\pi}
\frac{(m_1 m_2)^{3/2}}{(m_1 + m_2)^2 \epsilon^{2}}
\left( (\epsilon \Delta_{max} + E_i - E_f)^2 - (m_1 + m_2)^2 \right)^{5/2}
\left( \langle {\cal W}_{if} \rangle \omega/10 \right)
\,.
\nonumber \\ &&
\end{eqnarray}
Above a field threshold of ($m_1 + m_2 = 2 m_{\nu}$)
\begin{eqnarray}
&&
\frac{2m_e\omega m_{\nu}}{e\Delta_{max}} \sim
3 \,Vcm^{-1} \frac{m_{\nu}}{50 meV}
\frac{\omega }{GHz}\frac{400 eV}{\Delta_{max}}
\,,
\end{eqnarray}
the rate quickly rises to a rate of order,
\begin{eqnarray}
&&
w \approx 
\frac{G_F^2 \Delta_{max}^5 m_{\nu}}{240\pi}(\frac{eE_0}{m_e \omega})^3
\langle {\cal W}_{if} \rangle
\,,
\end{eqnarray}
taking $m_i = m_{\nu}$.
This rate is $m_{\nu}/\Delta_{max}$ times the rate much above the threshold.

A large factor by $(\Delta_{max}/m_{\nu})^5$ in these rates
is due to a larger momentum spread in sharply locarized
circular Rydberg states.

The critical field strength for the circular polarization
is a factor $\sqrt{2}$ larger, but the phase space is also different.

It is important to keep in mind that the coherence
should be maintained during the microwave irradiation.
If this is possible for a long time, one may expect a huge
growth factor $\langle {\cal W}_{if}\rangle $.
The ultimate bound on $\langle {\cal W}_{if}\rangle $
is derived by the unitarity argument in the following way
\cite{unitarity}.
The unitarity for time dependent process requires
\begin{eqnarray}
&&
\sum_f \int_{-\infty}^{T} dt' w_{fi}(t') \leq 1
\,.
\label{unitarity}
\end{eqnarray}
With a given time dependent rate $w_{fi}(t)$, this is
essentially a bound on the allowed time $T$.
It is appropriate to parametrize our weak process with
\begin{eqnarray}
&&
w_{fi}(t) = A \left((\frac{n_i}{n_f})^2 - 1 \right)^5 e^{\lambda \omega t}
\,,
\end{eqnarray}
due to the energy dependence of the weak transition
$n_i \rightarrow n_f$ by the neutrino pair emission.
The most important dependence is on the final state $n_f$,
as indicated. We ignored less important dependence of
$\lambda \,, A$ on $n_f$.
The requirement of unitarity eq.(\ref{unitarity}) is then
roughly
\begin{eqnarray}
&&
\frac{A}{\lambda \omega} e^{\lambda \omega T} \int_{n_0}^{n_f} dn_f 4n_f
\left((\frac{n_i}{n_f})^2 - 1 \right)^5 \leq 1
\,.
\end{eqnarray}
For estimate of $n_0$ we take the allowed lowest state
for the pair emission, $E_i - E_f > 2 m_{\nu}$.
\begin{eqnarray}
&&
\frac{\alpha^2 m_e}{2n_0^2} < 2 m_{\nu}
\,.
\end{eqnarray}
Taking as an order of magnitude estimate
$n_0 = \sqrt{\alpha^2 m_e/(4m_{\nu})}$ gives 
\begin{eqnarray}
&&
\frac{A}{\lambda \omega} e^{\lambda \omega t} \leq 
\frac{5 n_0^{10}}{2 n_i^{12}}
\sim
3 \times 10^{15} (\frac{20}{n_i})^{12}(\frac{50 meV}{m_{\nu}})^5
\,,
\end{eqnarray}
corresponding to the maximal rate,
\begin{eqnarray}
&&
\hspace*{-1cm}
A e^{\lambda T}\left((\frac{n_i}{n_f})^2 - 1 \right)^5
\leq 
3 \times 10^{5} s^{-1} \frac{\omega}{GHz}(\frac{50 meV}{m_{\nu}})^5
\left((\frac{n_i}{n_f})^2 - 1 \right)^5
\,,
\label{unitarity bound}
\end{eqnarray}
with $\lambda = 10$.

The actual rate would be very large,
much larger than the number given in eq.(\ref{rate estimate 0}), but
presumably less than the unitarity bound (\ref{unitarity bound}).
A practical rate might be limited by actual experimental
conditions such as the loss rate of coherence of
Rydberg atoms. 
Most relaxation time may be arranged to be much larger
than $10/ \omega \approx 10^{-8} s (GHz /\omega)$
such that the real rate may be close to the unitarity
bound.
There appears a real possibility of
measuring the neutrino pair emission process from circular
Rydberg states if the background rejection is successful.

It is beyond the scope of our present work to precisely 
locate and further exploit the wide band regions for the parametric
amplification of the neutrino pair emission.
Detailed study of this aspect will appear elsewhere.

\vspace{1cm}
\section{A few comments on experimental method}

\vspace{0.5cm}
\hspace*{0.5cm}
Clearly, one needs a systematic study, both theoretical and 
experimental, to implement our idea of the neutrino pair emission
from circular Rydberg atoms.
We shall be content here to make a few, rather trivial, comments
towards a more organized study.

Laser irradiated pair emission process is relatively
straightforward, hence we shall focuss on the microwave
irradiated Rydberg atoms.

\vspace{0.5cm}
\subsection{Key ideas for experimental success}

\hspace*{0.5cm}
One has to avoid the danger of disapperance of the intial Rydberg
states via ordinary radiative decay, since it might also be
enhanced by the wide band parametric resonance.
For this purpose, we may
use circular Rydberg states as the initial prepared state, 
for which radiative decays are very much suppressed 
except $n \rightarrow n-1$ E1 transition.
This E1 transition can be suppressed for instance by the cavity QED effect 
\cite{inhibited emission};
radiative decay mode within a cavity is modified by the
boundary effect, and if the wavelength $\lambda > 2d$, a size
of the cavity, the decay may be inhibited.
The inhibition prevails for all multiphoton transitions, since
all photons in this case satisfy the same condition $\lambda > 2d$
once one photon transition is inhibited.
One must however arrange experimental apparatus such that
the cavity does not interfere the microwave irradiation
\cite{sasao}.

Another important issue is how to unambiguously identify
the pair emission process.
Identification of final states is most important
for the determination of the threshold, and
this can be done by the field ionization technique \cite{rydberg atom}.
The threshold for the neutrino pair $\nu_i \nu_j$ appears at
\begin{eqnarray}
&&
\epsilon \Delta_{if} + E_i - E_f = m_i + m_j
\,,
\end{eqnarray}
with $\epsilon = eE_0/(m_e \omega)$.

The best way for unambiguous identification of
the weak process is to measure parity violating (PV) effects which
are absent in QED processes.
The simplest of this kind is to use circularly polarized photon beam
and to measure the difference of atomic transition rate
between $h = \pm 1$ polarization.
The asymmetry of the emitted photon distribution along the direction
of the polarization $\propto \vec{J}\cdot \vec{p}_{\gamma}$
is another  parity violating measurable.
Detailed theoretical study of this effect will appear elsewhere.

\vspace{0.5cm} 
\subsection{How to proceed}

\hspace*{0.5cm}
Since the small rate is a critical issue, an organized strategy of
experimental efforts is important.
We believe that the first step should be discovery of the neutrino pair
emission from excited atoms, and then one should steadily approach 
the smaller energy scale towards the pair emission threshold.
Presumably, at a few times twice the heaviest
neutrino mass of order $0.05 eV$, namely at energy $\approx 0.3 eV$,
one can hope to observe a signature of the difference between the Majorana 
and the Dirac neutrino.
The final step is a precision neutrino mass spectroscopy
along with measurement of mixing angles.

The use of atoms for neutrino physics is a new concept.
It is evident that many R and D are required for this project, but
one can initiate both experimental and theoretical efforts
by use of modest human and budgetary resources.
This is perhaps the ideal way towards a difficult
physics goal.

\vspace{1cm}
I would like to thank A. Fukumi, I. Nakano, H. Nanjo, and
N. Sasao for many stimulating discussions, in particular on
experimental feasibility of the ideas presented here.
N. Sasao has also pointed out a few mistakes in the original
version of this manuscript, which is much appreciated.
I also appreciate 
Y. Okabayashi for providing Figure 2 in the revised version
of this paper.

\vspace{1cm}
{\bf Note added}

After submission of this paper for publication,
we realized that laser irradiated pair emission
at lower thresholds $2m_1$ and $m_1 + m_2$ is useful
for detection of the relic cosmic neutrino of $1.9 K$
via the Pauli blocking effect.
See \cite{relic} for this observation.

\vspace{1cm}
{\bf Note added in proof}

After submission of this paper for publication
analysis based on the  optical Bloch equation
has been performed, and 
the greatest enhancement factor when 2 laser is irradiated,
adding another laser to $|I^* \rangle \leftrightarrow | I^{**} \rangle $ transition, has been obtained.
This work will be published elsewhere.

\vspace{1cm}
\section{Appendix A; Basic formulas of Majorana field}

\vspace{0.5cm} 
\subsection{Dirac and Majorana fields}
\hspace*{0.5cm} 
Using the representation of the Clifford algebra 
\begin{eqnarray}
&&
\gamma_{\alpha}\gamma_{\beta} + \gamma_{\beta}\gamma_{\alpha}
= 2 g_{\alpha\beta}
\,,
\end{eqnarray}
for $\gamma_{\alpha}$ (Greek $\alpha = 0\,, 1\,, 2\,, 3$ and 
Roman $i = 1\,, 2\,, 3$),
\begin{eqnarray}
&&
\gamma_0 =
\left(
\begin{array}{cc}
0 & 1 \\
1 & 0
\end{array}
\right)
\,, \hspace{0.5cm}
\gamma_i =
\left(
\begin{array}{cc}
0 & \sigma_i \\
- \sigma_i & 0
\end{array}
\right)
\,,
\\ &&
\gamma_0\gamma_i =
\left(
\begin{array}{cc}
- \sigma_i & 0 \\
0 & \sigma_i
\end{array}
\right)
\,, \hspace{0.5cm}
\gamma_5 = i \gamma_0\gamma_1\gamma_2\gamma_3
= 
\left(
\begin{array}{cc}
-1 & 0 \\
0 & 1
\end{array}
\right)
\,,
\\ &&
\hspace*{-1cm}
\psi = 
\left(
\begin{array}{c}
\varphi  \\
\chi
\end{array}
\right)
\,, \hspace{0.5cm}
\left(
\begin{array}{c}
\varphi  \\
0
\end{array}
\right) =
\frac{1}{2}(1 - \gamma_5)\psi 
\,, \hspace{0.5cm}
\left(
\begin{array}{c}
0  \\
\chi
\end{array}
\right) =
\frac{1}{2}(1 + \gamma_5)\psi
\,,
\end{eqnarray}
the Dirac equation
\begin{eqnarray}
&&
i \gamma_{\alpha}\partial^{\alpha} \psi - m \psi = 0
\,,
\end{eqnarray}
is written in the following 2-component form;
\begin{eqnarray}
&&
(i\partial_t - i\vec{\sigma}\cdot\vec{\nabla}) \varphi = m \chi
\,, \hspace{0.5cm}
(i\partial_t + i\vec{\sigma}\cdot\vec{\nabla}) \chi = m \varphi
\,.
\label{dirac eq 2}
\end{eqnarray}

The identification, $\chi = i\sigma_2 \varphi^*$,
in the Dirac equation gives the Majorana equation;
\begin{eqnarray}
&&
(i\partial_t - i\vec{\sigma}\cdot\vec{\nabla}) \varphi = i m 
\sigma_2 \varphi^*
\,,
\end{eqnarray}
with $m$ the neutrino mass.
The 2-component spinor $\varphi$ belongs to
$(1/2\,, 0)$ of the irreducible representation of the Lorentz group,
while $i\sigma_2 \varphi^*$ to $(0\,, 1/2$.

Explicit solution of helicity eigenstate is derived by
solving the helicity eigen-equation of eigenvalue $h=\pm 1$ ,
\begin{eqnarray}
&&
(\frac{\vec{\sigma}\cdot\vec{p}}{p} - h)
\left(
\begin{array}{c}
  a \\
  b
\end{array}
\right)
= 0
\,, \hspace{0.5cm}
\left(
\begin{array}{c}
  a \\
  b
\end{array}
\right)
= N
\left(
\begin{array}{c}
  p + hp_3 \\
  h(p_1 + ip_2)
\end{array}
\right)
\,.
\end{eqnarray}
The full plane-wave solution to the Majorana equation is then
\begin{eqnarray}
&&
\varphi(x) = 
e^{-ip\cdot x}N \xi 
- e^{ip\cdot x}\frac{E_p + hp}{m}N^* i\sigma_2 \xi^*
\,,
\label{plane-wave solution 1}
\\ &&
\xi = \xi(\vec{p}\,, h) =
\left(
\begin{array}{c}
  p + hp_3 \\
  h(p_1 + ip_2)
\end{array}
\right)
\,.
\end{eqnarray}
An equivalent form of solution is obtained by using 
\begin{eqnarray*}
&&
\frac{E_p + hp}{m} = \sqrt{\frac{E_p + hp}{E_p - hp}}
\,.
\end{eqnarray*}
When the helicity operator $- i\vec{\sigma}\cdot\vec{\nabla} /|\vec{\nabla}|$ 
is applied,
the first term gives the multiplicative factor $h$, while
the second gives the factor $-h$.
Thus, the consistent quantum interpretation
of particle annhihilation and anti-particle creation of two terms
is given by eq.(\ref{anti-particle creation}).

The free quantum Majorana field is described by
the Lagrangian density \cite{aithinson-hey} of
\begin{eqnarray}
&&
\hspace*{-1cm}
{\cal L}_{M} = \frac{1}{2}\left(
\left( \varphi^{\dagger} i\partial_0 \varphi
+ \varphi^{\dagger} i\vec{\sigma}\cdot\vec{\nabla} \varphi
\right)
+ (h.c.)
-
im \left(\varphi^{\dagger}\sigma_2 \varphi^* 
- \varphi^T \sigma_2 \varphi \right)
\right)
\,.
\end{eqnarray}
The Majorana particle
must be quantitzed according to the anti-commutation rule;
\begin{eqnarray}
&&
\{\varphi_{\alpha}(\vec{x}\,, t)\,, \varphi_{\beta}^{\dagger}
(\vec{y}\,, t) \}_+ = 
\delta^3 (\vec{x} - \vec{y})\delta_{\alpha \beta}
\,.
\label{anticommutation}
\end{eqnarray}

Mode decomposition in terms of plane-waves is given by
\begin{eqnarray}
&&
\hspace*{-1cm}
\varphi (x) =
\sum_{\vec{p}\,, h}
\left[ 
c(\vec{p}\,, h)u(\vec{p}\,, h)e^{-i p\cdot x} 
+ c^{\dagger}(\vec{p}\,, -h)
\sqrt{\frac{E_p + hp}{E_p - hp}}
(-i\sigma_2) 
u^* (\vec{p}\,,  h)e^{i p\cdot x}
\right]
\,,
\nonumber \\ &&
\end{eqnarray}
where
\begin{eqnarray}
&&
\{c(\vec{p}\,, h) \,, c^{\dagger}(\vec{p}'\,, h') \}_+ = 
\delta _{\vec{p} \,, \vec{p}'} \delta_{h h'}
\,,
\\ &&
\{c(\vec{p}\,, h) \,, c(\vec{p}'\,, h') \}_+ = 0
\,.
\end{eqnarray}
Relation of discrete and continuous momenta is given by
\begin{eqnarray}
&&
\delta _{\vec{p} \,, \vec{p}'} = \frac{(2\pi)^3}{V} \delta^{(3)}
(\vec{p} - \vec{p}')
\,, \hspace{0.5cm}
\sum_{\vec{p}} f_{\vec{p}}= V \int \frac{d^3 p}{(2\pi)^3}f(\vec{p})
\,,
\end{eqnarray}
with $V$ the volume of the normalization box.

The normalization of 2-spinor
consistent with canonical anti-commutation eq.(\ref{anticommutation})
is derived as follows.
Computing the anti-cummutator with the plane-wave mode decomposition
gives
\begin{eqnarray}
&&
\hspace*{1cm}
\{\varphi_{\alpha}(\vec{x}\,, t)\,, \varphi_{\beta}^{\dagger}
(\vec{y}\,, t) \}_+ = 
\nonumber \\ &&
\sum_{\vec{p}\,, h}
\left[
e^{i\vec{p}\cdot (\vec{x} - \vec{y})}
|N(\vec{p}\,, h)|^2 (p + hp_3)(p + h \vec{\sigma}\cdot\vec{p})_{\alpha\beta}
\right.
\nonumber \\ &&
\left.
+
e^{-i\vec{p}\cdot (\vec{x} - \vec{y})}
\frac{E_p + hp}{E_p - hp}
|N(\vec{p}\,, h)|^2 (p + hp_3)(p - h \vec{\sigma}\cdot\vec{p})_{\alpha\beta}
\right]
\,.
\end{eqnarray}
We have used the relation,
\begin{eqnarray}
&&
\xi(\vec{p}\,, h) \xi^{\dagger}(\vec{p}\,, h)
= (p + hp_3)(p + h \vec{\sigma}\cdot\vec{p})
\,.
\end{eqnarray}
The correct anti-commutation relation thus requires
\begin{eqnarray}
&&
N(\vec{p}\,, h) = \frac{1}{2}\sqrt{\frac{E_p - hp}{pE_p (p + hp_3)}}
\,.
\end{eqnarray}
Useful relations on the normalized wave function,
\begin{eqnarray}
&&
u(\vec{p}\,, h) = \frac{1}{2}\sqrt{\frac{E_p - hp}{pE_p (p + hp_3)}}
\left(
\begin{array}{c}
  p + hp_3 \\
  h(p_1 + ip_2)
\end{array}
\right)
\,,
\end{eqnarray}
are
\begin{eqnarray}
&&
u(\vec{p}\,, h)u^{\dagger}(\vec{p}\,, h) = \frac{1}{4}
(1 - \frac{hp}{E_p})(1 + h\frac{\vec{\sigma}\cdot\vec{p}}{p})
\,,
\\ &&
u^{\dagger}(\vec{p}\,, h)u(\vec{p}\,, h) = \frac{1}{2}
(1 - \frac{hp}{E_p})
\,.
\end{eqnarray}

In quantum field theory it is important to identify
Hamiltonian, momentum, and propagator. They are given for the Majorana field,
\begin{eqnarray}
&&
{\cal H}(\vec{x})
= -i \varphi^{\dagger}\vec{\sigma}\cdot\vec{\nabla}\varphi  + (h.c.)
+ \frac{im}{2}
(\varphi^{\dagger} \sigma_2 \varphi^* - \varphi^T \sigma_2 \varphi )
\,,
\\ &&
H = \int d^3 x {\cal H}(\vec{x})
= \sum_{\vec{p}\,, h} \,
E_p c^{\dagger}(\vec{p}\,, h)\,c(\vec{p}\,, h) 
\,,
\label{field hamiltonian}
\\ &&
{\cal P} = \int d^3 x \varphi^{\dagger}(\vec{x})
(-i \vec{\nabla})\varphi(\vec{x})
= \sum_{\vec{p}\,, h} \, \vec{p}c^{\dagger}(\vec{p}\,, h)\,c(\vec{p}\,, h) 
\,,
\label{field momentum}
\\ &&
\langle 0 | T\left( \varphi(y) \varphi^{\dagger}(x) \right) | 0 \rangle
= - \sigma \cdot \partial
\int \frac{d^4 p}{(2\pi)^{4}} 
\frac{e^{i p \cdot (y - x)}}{p^2 - m^2 + i\epsilon}
\,,
\\ &&
\langle 0 | T\left( \varphi(y) \varphi(x) \right) | 0 \rangle
= im\sigma_2
\int \frac{d^4 p}{(2\pi)^{4}} 
\frac{e^{i p \cdot (y - x)}}{p^2 - m^2 + i\epsilon}
\,.
\label{majorana mass propagation}
\end{eqnarray}
The energy (\ref{field hamiltonian}) and the momentum (\ref{field momentum})
formulas of the Majorana field establish
the corecctness of identification of $c^{\dagger}(\vec{p}\,, h)$ and 
$c(\vec{p}\,, h)$ as particle creation and annihilation operators.
The second form of the propagator (\ref{majorana mass propagation}) is characteristic of
Majorana field that does not conserve the fermion number,
and does vanish for the Dirac field.

\vspace{1cm}
\section{Appendix B; Weak perturbative process for Rydberg atoms under 
microwave irradiation}

\vspace{0.5cm} 
\hspace*{0.5cm}
Consider a Hamiltonian system $H_0$ perturbed by
2 types of generally time-dependent interaction,
one of which with $H_0 + V_A(t)$ is treated as solvable,
and $V_B$ as very weak;
\begin{eqnarray}
&&
H = H_0 + V_A(t) + V_B
\,, \hspace{0.5cm}
H_0 = \frac{\vec{p}^2}{2m} + V_C(\vec{x}) \,,
\\ &&
(i \partial_t - H_0 - V_A(t)) |A_n(t) \rangle = 0 \,,
\label{unperturbed s-eq}
\\ &&
\hspace*{-1cm}
|A_n(t) \rangle = 
U_A(t\,, t_0)|A_n(t_0) \rangle
\,, \hspace{0.5cm}
U_A(t\,, t_0) = \exp[- i \int_{t_0}^t dt_1 (H_0 + V_A(t_1)\,)] \,,
\nonumber \\ &&
\\ &&
\langle A_n(t) | =
\langle A_n(t_0) | U_A(t_0\,, t )
= \langle A_n(t_0') | U_A(t_0'\,, t ) \,.
\end{eqnarray}
In the interaction picture, the weak vertex is
\begin{eqnarray}
&&
V_B(t) = U_A^{-1}(t\,, t_0)V_BU_A(t\,, t_0) \,.
\end{eqnarray}
The transition S-matrix $S$
is computed using solutions $|A_n(t) \rangle$ of (\ref{unperturbed s-eq}), from
\begin{eqnarray}
&&
\hspace*{-1cm}
\langle A_f(t_0) |V_B(t_1)|A_i(t_0) \rangle
= \langle A_f(t_0') |U_A^{-1}(t_1\,, t_0')V_BU_A(t_1\,, t_0)|A_i(t_0) \rangle \,.
\end{eqnarray}
By taking the time limit of $t_0' \rightarrow \infty $ and 
$t_0 \rightarrow - \infty$,
the transition matrix element is to lowest order of $V_B$,
\begin{eqnarray}
&&
\hspace*{-1cm}
(S - 1)_{fi} \sim - i \int_{- \infty}^{\infty}\, dt_1\,
\langle A_f(\infty)| U_A^{-1}(t_1\,, \infty)V_B
U_A(t_1\,, -\infty) | A_i(-\infty)\rangle 
\,.
\nonumber \\ &&
\end{eqnarray}
Expand states at time $= \pm \infty$ 
in terms of a complete set of momentum eigenstates;
\begin{eqnarray}
&&
|A_n(-\infty) \rangle = 
\int \frac{d^3 p\,|\vec{p} \rangle \langle \vec{p} |A_n(-\infty) \rangle}
{(2\pi)^{3/2}} 
\,.
\\ &&
\hspace*{-1cm}
(S - 1)_{fi} \sim - i \int_{- \infty}^{\infty}\, dt_1\,
\int \frac{d^3 p_f\,\langle A_f(\infty)| 
\vec{p}_f \rangle \langle \vec{p}_f | }{(2\pi)^{3/2}}
U_A(\infty \,,t_1)V_BU_A(t_1\,, - \infty)
\nonumber \\ &&
\hspace*{0.5cm}
\times
\int \frac{d^3 p_i\,|\vec{p}_i \rangle \langle \vec{p}_i 
| A_i(- \infty)\rangle }
{(2\pi)^{3/2}} 
\,.
\label{s-matrix}
\end{eqnarray}
We shall need to compute the transition matrix element
for momentum eigenstates;
\begin{eqnarray}
&&
 \langle \vec{p}_f |U_A(\infty \,,t_1)V_B 
U_A(t_1\,, - \infty)|\vec{p}_i \rangle 
\,,
\end{eqnarray}
sandwitched between momentum state wave functions of
\begin{eqnarray}
&&
\langle A_f(\infty)|\vec{p}_f \rangle
\,, \hspace{0.5cm}
\langle \vec{p}_i | A_i(-\infty)\rangle
\,.
\end{eqnarray}

States here
\begin{eqnarray}
&&
U_A(t\,, - \infty)| \vec{p}_i \rangle
\,, \hspace{0.5cm}
(\vec{p}_f |U_A(\infty \,,t))^{\dagger} 
= U_A(t\,,  \infty)| \vec{p}_f \rangle
\,,
\end{eqnarray}
are solutions of the Schroedinger equation,
eq.(\ref{unperturbed s-eq}).

These momentum state wave functions for
H-atom are given in \cite{bs}.
For instance, those of circular states of
$l = |m| = n -1$ are
\begin{eqnarray}
&&
\psi_{nn-1m}(\vec{p}) \propto
e^{im\varphi}(1 - \cos^2 \theta)^{(n-1)/2}
\frac{(np)^{n-1}}{(\,(np)^2 + p_0^2)^{n+1}} \,,
\end{eqnarray}
with $m = \pm (n-1)$.
For general states, using the atomic unit of $p_0 = 1/a_B$
\begin{eqnarray}
&&
\psi_{nlm}(\vec{p}) = F_{nl}(p) Y_{lm}(\theta \,, \varphi)
\,,
\\ &&
F_{nl}(p)
\propto
\frac{(np)^l}{(\,(np)^2 + 1)^{l+ 2}}
C_{n-l-1}^{l+1}(\frac{n^2 p^2 - 1}{n^2 p^2 + 1})
\,,
\end{eqnarray}
where $C_{N}^{\nu}(x)$ is the Gegenbauer polynomial (of orfer $N$).

Both states, $U_A(t\,, - \infty)|A_i(-\infty) \rangle $ and
$U_A(\infty \,, t)|A_f(\infty) \rangle$, contain the Coulomb and microwave field effects.
Keldysh \cite{keldysh} neglects the Coulomb interaction.
According to Reiss \cite{reiss}, this is justified if 
\begin{eqnarray}
&&
r = \frac{\omega^2 a_B}{m_e} \times \frac{E^2}{\omega^4}
\,,
\end{eqnarray}
is small.
($\frac{E^2}{\omega^4}$ is the number of modes in the field).
This reduces to
\begin{eqnarray}
&&
r \leq \frac{E_c^2}{\alpha \omega^2 m_e^2} 
\,,
\end{eqnarray}
in our problem.
$E_c$ is the maximum field allowd below
ionization, given by
\begin{eqnarray}
&&
eE_c = \frac{\alpha}{(n^2 a_B)^2}
\,.
\end{eqnarray}
Thus, the above parameter 
\begin{eqnarray}
&&
r \leq (\alpha ^2\frac{m_e}{\omega})^2 \frac{1}{n^8}
\sim 4.3 \times 10^5
(\frac{GHz}{\omega})^2 (\frac{10}{n})^8
\,.
\end{eqnarray}
This parameter is small if
\begin{eqnarray}
&&
\omega \leq 2.1 \times 10^6 GHz n^{-4}
\,,
\end{eqnarray}
and 
\begin{eqnarray}
&&
E \leq E_C = \alpha^{5/2} m_e^2 n^{-4}
\sim 5.2 \times 10^9 V cm^{-1}
\,.
\end{eqnarray}

Thus, we arrive at 
\begin{eqnarray}
&&
\hspace*{-1cm}
U_A(t\,, - \infty)| \vec{p} \rangle
\sim \exp [-i \int_{- \infty)}^t dt_1 (\frac{\vec{p}^2}{2m} + 
\frac{\vec{p}\cdot\vec{A}(t_1)}{m} + \frac{e^2\vec{A}^2(t_1)}{2m} )]
| \vec{p} \rangle
\,,
\label{volkov solution}
\end{eqnarray}
which is the Volkov solution
\cite{atomic physics}, the exact plane-wave solution under a periodic
field.
Using an explicit form of $\vec{A}(t)$,
\begin{eqnarray}
&&
\int_{t_0)}^t dt_1 \vec{A}(t_1) =
\frac{\vec{E}_0}{\omega}\sin (\omega t) + ({\rm const.})
\,.
\end{eqnarray}

Using the Volkov state, (neglecting irrelevant constants), the matrix element
is
\begin{eqnarray}
&&
 \langle \vec{p}_f |U_A(\infty \,,t)V_B U_A(t\,, - \infty)|\vec{p}_i \rangle
\nonumber \\ &&
\sim 
\exp [-i \left(\frac{\vec{p}_i^2 - \vec{p}_f^2}{2m}\,t + 
\frac{(\vec{p}_i - \vec{p}_f)\cdot\vec{E}_0}{m \omega^2}\sin (\omega t)\,\right)]
\langle \vec{p}_f |V_B |\vec{p}_i \rangle
\,.
\label{volkov evolution}
\end{eqnarray}
Time dependence of the relevant matrix element of
the integrand of (\ref{s-matrix})
\begin{eqnarray}
&&
\langle A_f(\infty)|\vec{p}_f \rangle
\langle \vec{p}_f |U_A(\infty \,,t)V_B U_A(t\,, - \infty)|\vec{p}_i \rangle
\langle \vec{p}_i | A_i(-\infty)\rangle
\,,
\end{eqnarray}
is in eq.(\ref{volkov evolution}).

High frequency Floquet theory (HFFT) of \cite{hfft}
that predicts dressed Coulomb potential caused by
averaged field irradiation appears irrelevant in
our paremeter range of field strength and frequency.



\begin{thebibliography}{99}
\bibitem{fukumi}
The author should like to thank A. Fukumi for
preparing Figure 1.
\bibitem{unitarity}
Any complete theory satisfys the unitarity automatically.
Our treatment of the microwave irradiation as externally given
is however incomplete.
The argument presented here, in particular
the inequality (\ref{unitarity}), is the limit that must be
satisfied by any incomplete theory that treats 
the periodic external field as in our model.
\bibitem{sasao}
The author should like to thank N. Sasao for
a comment related to this problem.
\end{thebibliography}
\end{document}